\documentclass{article}
\pdfoutput=1 % if your are submitting a pdflatex (i.e. if you have
\usepackage{jheppub} % for details on the use of the package, please
\usepackage[T1]{fontenc} % if needed
\usepackage[utf8]{inputenc}
\usepackage{amsmath}
\usepackage{amssymb}
\usepackage{mathtools}
\usepackage[dvipsnames]{xcolor}
\usepackage{hyperref}
\usepackage{graphicx}
\usepackage{pgfplots}
\usepackage{subfig}
\usepackage{cleveref}
\usepackage{csquotes}
\usepackage{ulem}
\usepackage[boxed, linesnumbered]{algorithm2e} % For algorithms
\SetKwComment{comment}{}{}
\SetCommentSty{text}
\SetNlSty{texttt}{[}{]}	
\DeclareUnicodeCharacter{2212}{-} % Annoying workaround since
\newcommand{\ve}[1]{\bs{#1}} %How we write notation for vectors
\newcommand{\bs}[1]{\boldsymbol{#1}} %Not sure what this is, possibly the same as above?

\renewcommand{\d}{\mathrm{d}}

\newcommand{\CL}{CL}
\newcommand{\final}[1]{{#1}}

\let\oldnl\nl% Store \nl in \oldnl
\newcommand{\nonl}{\renewcommand{\nl}{\let\nl\oldnl}}

\title{Fast and robust Bayesian Inference using Gaussian Processes with \texttt{GPry}}
\author[a]{Jonas El Gammal,}
\author[b]{Nils Sch\"oneberg,}
\author[c,d]{Jes\'{u}s Torrado,}
\author[e]{and Christian Fidler}

\affiliation[a]{Department of Mathematics and Physics, University of Stavanger, NO-4036 Stavanger, Norway}
\affiliation[b]{Institut de Ci\`encies del Cosmos, Universitat de Barcelona, Mart\'{\i} i Franqu\`es 1, Barcelona E08028, Spain}
\affiliation[c]{Dipartimento di Fisica e Astronomia \enquote{G. Galilei}, Universit\`a degli Studi di Padova, Via Marzolo 8, I-35131 Padova, Italy}
\affiliation[d]{INFN, Sezione di Padova, Via Marzolo 8, I-35131 Padova, Italy}
\affiliation[e]{Institute for Theoretical Particle Physics and Cosmology (TTK), \\ RWTH Aachen University, D-52056 Aachen, Germany.}

\date{October 2021}
\emailAdd{jonas.e.elgammal@uis.no}
\emailAdd{nils.science@gmail.com}
\emailAdd{jesus.torrado@pd.infn.it}
\emailAdd{fidler@physik.rwth-aachen.de}

\begin{document}
\normalem

\abstract{
We present the \texttt{GPry} algorithm for fast Bayesian inference of general (non-Gaussian) posteriors with a moderate number of parameters. \texttt{GPry} does not need any pre-training, special hardware such as GPUs, and is intended as a drop-in replacement for traditional Monte Carlo methods for Bayesian inference. Our algorithm is based on generating a Gaussian Process surrogate model of the log-posterior, aided by a Support Vector Machine classifier that excludes extreme or non-finite values. An active learning scheme allows us to reduce the number of required posterior evaluations by two orders of magnitude compared to traditional Monte Carlo inference. Our algorithm allows for parallel evaluations of the posterior at optimal locations, further reducing wall-clock times. We significantly improve performance using properties of the posterior in our active learning scheme and for the definition of the GP prior. In particular we account for the expected dynamical range of the posterior in different dimensionalities. We test our model against a number of synthetic and cosmological examples. \texttt{GPry} outperforms traditional Monte Carlo methods when the evaluation time of the likelihood (or the calculation of theoretical observables) is of the order of seconds; for evaluation times of over a minute it can perform inference in days that would take months using traditional methods. \texttt{GPry} is distributed as an open source Python package (\texttt{pip install gpry}) and can also be found at  \url{https://github.com/jonaselgammal/GPry}.}

\maketitle

\section{Introduction}

% Traditional MC Bayesian inference
One of the fundamental tools of science is the comparison of observations with theory. In many situations, this involves inference on the parameters of a model (or on models themselves) given some observed data. This is often realised using Bayesian statistics, where one synthesises the probability of some data having been acquired into a \emph{likelihood function}, assumes some \emph{a priori} distribution for the model parameters, and samples from the product of both (proportional to the so-called \emph{posterior}) using Monte Carlo methods, the most common ones in Cosmology being based on Markov Chain Monte Carlo \cite{mcmc_sampler_2, mcmc_sampler_1, Foreman_Mackey_2013, Akeret_2013} or Nested Sampling \cite{skilling_nested_sampling,multinest_1,multinest_2,multinest_3,polychord_1,polychord_2,Higson_2018,Speagle_2020}.

% Traditional MC becoming unfeasible
The new era of cosmological surveys will produce data in rapidly increasing amount and quality \cite{Feigelson_2021,https://doi.org/10.48550/arxiv.2110.10074}. This will in turn raise the computational costs of traditional Monte Carlo pipelines: data quality will call for an increase in the precision of theoretical computations of the observables that are compared against the data (e.g. including physical effects that could have been so far neglected), and likelihood computations will involve operations on ever larger data vectors. This can and will eventually result in traditional Bayesian inference becoming prohibitively slow, further increasing the already damaging carbon footprint of scientific computations in computer clusters \cite{https://doi.org/10.48550/arxiv.1912.05834,Portegies_Zwart_2020}. In order to keep being able to exploit cosmological data for parameter inference, we need to develop more advanced algorithms that significantly reduce the computational costs of performing inference, and machine-learning based methods are one of the most promising tools for that.

% Solution: emulators
So far, a number of different solutions have been proposed. A family of them focus on substituting the theoretical computation of observables (or intermediate quantities to arrive at them) by appropriately-trained, usually Neural Network-based, \emph{emulators} that cheaply map the theoretical parameters onto the space of vectors of observables. For applications to Cosmology and Astrophysics, see e.g.\ \cite{Kaplinghat:2002mh,Jimenez:2004ct,10.1111/j.1745-3933.2006.00276.x,Auld:2007qz,Albers:2019rzt,Manrique-Yus:2019hqc,Mootoovaloo:2020ott,Nygaard:2022wri,Donald-McCann:2021nxc,Donald-McCann:2022pac,Bonici:2022xlo,Mootoovaloo:2021rot,Gunther:2022pto,10.1093/mnras/stac064,https://doi.org/10.48550/arxiv.2203.05583, https://doi.org/10.48550/arxiv.2008.12932, Chianese_2020, Lanusse_2021, Rogers_2019,
McClintock_2019, Ho_2021, Khan_2021, Moore_2014, https://doi.org/10.48550/arxiv.2012.05472, Manrique_Yus_2019, Bird_2019, https://doi.org/10.48550/arxiv.2205.15726}. These methods are robust in the sense that they are guaranteed to reproduce the true posterior distribution, as long as the emulator is properly trained, which is easy to check a posteriori. Unfortunately their utility is limited by the need to retrain them whenever the theoretical model under investigation is varied. Additionally, as experiments become ever more precise, in order to achieve sufficient accuracy a larger number of systematic effects needs to be accounted for, which requires ever more costly experimental likelihoods, which cannot be easily accelerated by emulators.

% Solution: likelihood-free
Another proposed solution are simulation-based \textit{likelihood-free} approaches, inspired by Approximate Bayesian Computation, but accelerated by Neural Networks \cite{Cranmer_2020,https://doi.org/10.48550/arxiv.2101.04653}. There, Neural Networks are used to learn a mapping between sets of model parameters and their corresponding simulated data, so that they can automatically extract features, marginalise over nuisance parameters, learn a likelihood function, or ultimately produce a posterior distribution of the model parameters when fed real experimental data. Recent development and applications in Cosmology and Astrophysics can be found in \cite{https://doi.org/10.48550/arxiv.2010.12931, Alsing_2019, https://doi.org/10.48550/arxiv.2011.13951, Hermans_2021, Gerardi_2021, Huppenkothen_2021, https://doi.org/10.48550/arxiv.2105.12024, Zhang_2021, https://doi.org/10.48550/arxiv.2211.00723,https://doi.org/10.48550/arxiv.2208.00134, https://doi.org/10.48550/arxiv.2203.06124, https://doi.org/10.48550/arxiv.2204.04491}. The claimed advantages are that they may discover or take into account features in the data that are not captured by summary statistics or observables, and the lack of need to formulate a likelihood, which can be complex or prohibitively expensive in some cases. On the other hand, they tend to require expensive training and the reusability of the trained networks is limited when considering model extensions. The need to accurately account for modelling uncertainty and possible biases has also been highlighted recently \cite{Grand_n_2022, https://doi.org/10.48550/arxiv.2207.08435}.

% Our solution: GP emulation of the posterior

The solution presented in this work differs from the previous ones in that it retains the full computation of the observable and data likelihood, but minimises the number of points in the parameter space where this full pipeline needs to be computed; it uses these points to create a model of the posterior, and to iteratively predict the next optimal evaluation locations. For the emulation of the posterior we use Gaussian Processes (GP) \cite{gpml}, which have a small number of hyperparameters that are easily interpretable in terms of properties of the posterior, and thus make it easier to incorporate prior information about its functional form. Furthermore, due to their simplicity, Gaussian Processes generally require smaller training sets than for example Neural Networks. We combine the GP model of the posterior with a support vector machine (SVM) \cite{Murphy2012, Cortes1995} to restrict the parameter space to a region of reasonable posterior values.

Our approach expands on previous work applying Bayesian quadrature and active sampling to statistical inference \cite{Henning_inference,ALoMEuBQ}, which we improve upon by incorporating the expected scaling of the log-posterior with dimensionality, the definition of a cheap and consistent convergence criterion and the treatment of extreme log-posterior values with an SVM classifier. A previous attempt at a similar approach to inference in Cosmology with a GP surrogate of the posterior can be found in \cite{paper_1}, and in the context of emulator-training in \cite{lyman_alpha}.

% The final product
The result of our work is the development of the \texttt{GPry} algorithm. An open source implementation is available as a Python package (\texttt{pip install gpry}) and at \url{https://github.com/jonaselgammal/GPry}. \texttt{GPry} does not need any pre-training or parameter tuning, so it can be used as a \emph{drop-in} replacement for traditional Monte Carlo algorithms. Unlike neural networks it also does not require any specialised hardware such as GPUs. As we will show, it allows for accurate and fast emulation of posteriors for moderate dimensionalities, including non-Gaussian distributions, by using just a few hundred or thousand evaluations of the posterior distribution. Especially when individual likelihood evaluations are computationally expensive, this can result in large, speedups of typically two orders of magnitude.

This paper is structured as follows: in \Cref{sec:basics} we review the basic concepts and useful notation. We continue in \Cref{sec:code} presenting the modelling choices involved in the construction of the GP surrogate model. The learning strategy for acquiring new sampling locations as well as a criterion for deciding on convergence are discussed in \Cref{sec:strategy}. In \Cref{sec:full_algorithm} we put together all the pieces and present the full algorithm, and comment on its general performance. We discuss the performance of \texttt{Gpry} on different synthetic and cosmological problems in \Cref{sec:examples}, and we present our conclusions and discuss possible future development in \Cref{sec:conclusions}. \Cref{app:threshold} is dedicated to discussing the inclusion of prior information on the dynamical range of the posterior into the surrogate model at different stages of the algorithm.

\section{Basic concepts}\label{sec:basics}

In order to establish a consistent notation and a deeper understanding of the underlying concepts, we quickly summarize some of the theory, which we are going to use in the detailed description of \Cref{sec:code}.

\subsection{Bayesian inference of model parameters}\label{ssec:bayesian_inference}

A usual Bayesian inference problem is that of estimating the probability distribution $p(\mathcal{M}(\ve x)|\mathcal{D})$ of the parameters $\ve x$ of a model $\mathcal{M}$ given some experimental data $\mathcal{D}$, also known as \emph{posterior}. Following Bayes' theorem, this is proportional to the product of the \emph{likelihood} $p(\mathcal{D}|\mathcal{M}(\ve x))$ (the probability of $\mathcal{D}$ having being measured given the model with these parameter values), and the \emph{prior} probability of the parameter values $\ve x$ given the model, $p(\ve x|\mathcal{M})$, assigned before (or idependently of) the experiment that measured $\mathcal{D}$.\footnote{The missing proportionality constant is the inverse of the \emph{evidence} $p(\mathcal{D}|\mathcal{M})$, which can usually be ignored in parameter estimation and will hence be omitted in all subsequent calculations. Note though, that the evidence is important when performing model selection.} Fixing the model $\mathcal{M}$ and the data $\mathcal{D}$, we can drop their explicit dependence to simplify notation. With that Bayes' theorem reads
\begin{equation}\label{eq:post}
    p(\ve x) \propto \mathcal{L}(\ve x) \pi(\ve x)\,,
\end{equation}
where $p(\ve x)$ is the posterior, $\mathcal{L}(\ve x)$ the likelihood, and $\pi(\ve x)$ the prior.

In Cosmology, likelihoods are typically provided by experimental collaborations, are generally non-analytic, or analytic but non-differentiable, and usually also costly to evaluate. Even when they are well-behaved, they sometimes depend on cosmological quantities whose computation in terms of the parameters to be inferred has the same undesirable properties. In these cases, the targeted solution to the inference problem is obtaining a Monte Carlo sample of the posterior, often using MCMC- or nested-sampling-based methods.

This work focuses on reducing the number of evaluations of the posterior (and thus the likelihood) needed to solve the inference problem. We do that by creating a surrogate model of the posterior using a Gaussian Process, and developing an active learning algorithm that decides sequentially on a small optimal set of parameter values where to evaluate the true likelihood, so that the surrogate model is accurate enough. One can then e.g.\ extract the usual Monte Carlo sample from the resulting surrogate model of the posterior (which, as a bonus, is differentiable) at a very low computational cost.

If the goal is to obtain 1D/2D posteriors (and their corresponding \CL{}s) from the GP, one could wonder if there would be alternative efficient methods of computing the required marginalization integrals. However, generally the integrals involved are not solvable analytically and due to the high dimensionality of these integrals in most applications, the most efficient ways of computing them numerically are usually Monte-Carlo methods. We discuss the computational costs of this choice in \Cref{ssec:timecosts}.

\subsection{Gaussian Processes}\label{ssec:gp}

We briefly present the relevant GP notation and formulae that we will need for this work. For a more thorough review, see \cite{gpml}.

Gaussian Processes are useful to emulate a sufficiently smooth\footnote{Here, "sufficiently smooth" refers to an underlying function which $n$-times continuously differentiable where $n\geq1$. The function may still have some statistical or numerical noise added on top of it.} function $f(\ve{x})$ at an arbitrary point $\ve{x}$ (within a certain domain) given a set of sampling locations $\ve{X}=\left\{\ve{x}_1,\ldots,\ve{x}_{N_s}\right\}$ and their corresponding function values $y_i = f(\ve{x}_i)$ for $i=1\,\ldots N_s$. This last equation can be abbreviated to $\ve{y} = \ve{f}(\ve{X})$ (notice the bold symbol for $\ve{y}$ and $\ve{f}$ and the dependence on $\ve{X}$) following the usual notation in GP literature, where the number of samples is treated as an additional vector space of dimension $N_s$, with components denoted by a subscript. This means that $\ve{X}$ becomes a $N_s \times N_d$ matrix, where $N_d$ is the dimensionality of the parameter space. This way, we write for a scalar function $s(\ve{x})$ evaluated at the $N_s$ different sampling locations the vector $\ve{s}(\ve{X})$ with components $[\ve{s}(\ve{X})]_i = s(\ve{X}_i)$, and similarly for scalar functions of two arguments the tensor $\ve{s}(\ve{X},\ve{X})$ with components $[\ve{s}(\ve{X},\ve{X})]_{ij} = s(\ve{X}_i,\ve{X}_j)$.

A Gaussian Process posits that the function $f(\ve{x})$ in question is a random draw from a family of functions, informed by the sampling locations. For a given position $\ve{x}$ such random draw of a function $f(\ve{x})$ is assumed to be Gaussian-distributed (hence the name) around a mean function $m(\ve{x})$ with a covariance between the functional value at two different points given by some function $k(\ve{x},\ve{x}^\prime)$, often called the \emph{kernel} of the GP.
\begin{equation}\label{unconditional}
    \widehat{f} \sim \mathcal{GP}(m,k)  \qquad \Leftrightarrow \qquad \widehat{f}(\ve{x}) \sim \mathcal{N}(m(\ve{x}), k(\ve{x},\ve{x}))\,,
\end{equation}
where $\widehat{f}$ denotes a random function draw from the GP and $\sim$ means \enquote{is distributed according to}. As a multivariate Gaussian distribution, the GP is completely defined by its mean and kernel functions. Their precise choice only aids in faster and more predictive emulation, but they do not in general restrict the shape of the functions being modeled, which can be complete arbitrary as long as the kernel function fulfills a number of weak conditions \cite{universal_kernels} (that all kernels considered in this work do). Importantly, while the \emph{correlation} of the function value at two points is assumed to be Gaussian, this neither means that the function is itself assumed to be Gaussian, nor that the mean of the family of functions is presumed to be Gaussian.

We usually restrict the GP so that it agrees with the given set of sampling locations for all draws, $\ve{\widehat{f}}(\ve{X})\stackrel{!}{=}\ve{f}(\ve{X})=\ve{y}$, sometimes up to some uncorrelated Gaussian noise. This information modifies the value of the drawn function's \emph{predictions} $\ve{\widehat{f}}(\ve{X}_*)$ away from the sampled values $\ve{X}$. The joint distribution for sampled and non-sampled locations is
\begin{align}\label{joint_distribution}
\begin{bmatrix}
\ve{\widehat{f}}(\ve{X}) \\ \ve{\widehat{f}}(\ve{X}_*)
\end{bmatrix}\sim \mathcal{N}\left(\begin{bmatrix}
\ve{m}(\ve{X}) \\ \ve{m}(\ve{X}_*)
\end{bmatrix}, \begin{bmatrix}
\ve{k}(\ve{X},\ve{X}) & \ve{k}(\ve{X},\ve{X}_*) \\ 
\ve{k}(\ve{X}_*,\ve{X}) & \ve{k}(\ve{X}_*,\ve{X}_*)
\end{bmatrix}\right)
\end{align}
This defines the \emph{conditional} probability for the predictions $\ve{\widehat{f}}(\ve{X}_*)$ given the observations $(\ve{X},\ve{y})$ as
\begin{align}\label{conditional}
\widehat{\ve{f}}|_{\ve{f}(\ve{X})=\ve{y}} \sim \mathcal{GP}(\ve{\mu},\ve{\Sigma}) \qquad \Leftrightarrow \qquad \ve{\widehat{f}}(\ve{X}_*) |_{\ve{f}(\ve{X})=\ve{y}}\sim \mathcal{N}(\ve{\mu}(\ve{X}_*),\ve{\Sigma}(\ve{X}_*))~.
\end{align}
with mean vector and covariance matrix
\begin{align}
\ve{\mu}(\ve{X}_*) &= \ve{m}(\ve{X}_*) + \ve{k}(\ve{X}_*, \ve{X}) \ve{k}(\ve{X},\ve{X})^{-1} \left[\ve{y}-\ve{m}(\ve{X})\right]~,\label{mean_gp}\\
\ve{\Sigma}(\ve{X}_*) &= \ve{k}(\ve{X}_*,\ve{X}_*) - \ve{k}(\ve{X}_*,\ve{X})\ve{k}(\ve{X},\ve{X})^{-1}\ve{k}(\ve{X},\ve{X}_*)~.\label{cov_gp}
\end{align}
This conditioned GP for new sample predictions is then called the \textit{posterior GP}.
Comparing \Cref{unconditional,conditional} we notice that the drawn samples $\widehat{\ve{f}}$ differ between the unconditioned and the conditioned GP, because the latter includes the additional information from the sampling locations.
The algorithm described in \Cref{sec:full_algorithm} will sequentially add new samples to the GP. These will incorporated by \emph{updating} the mean and covariance of this conditioned GP ($\mu$,$\Sigma$) using sequentially enlarged sample sets $(\ve{X},\ve{y})$.

From here on, we will use the scalar versions of \Cref{mean_gp,cov_gp} evaluated at an arbitrary single location $\ve{x}$ as $\mu(\ve{x})$ and $\Sigma(\ve{x})$, as well as $\sigma(\ve{x}) = \sqrt{\Sigma(\ve{x})}$ as the uncertainty of the GP at a location~$\ve{x}$ to simplify notation, implicitly assuming it has been conditioned on the samples $\ve{X}$. As is standard in the literature (and as discussed without loss of modeling power for the GP), we will assume a zero-mean function $m(\ve{x})=0$ in all cases.

Kernel functions are usually chosen from a particular family of functions (such as squared exponentials, \emph{Mat\'ern} kernels,~\ldots),\footnote{The kernel function is typically chosen according to the differentiability and smoothness of the given target function, see also \cref{ssec:kernel} for more details.} parameterized by some \emph{hyperparameters} $\theta$. Their value is commonly chosen so that they maximize the likelihood that the GP would have produced the given sampled values $\ve{y}$ at the sampled locations $\ve{X}$. In practice, one marginalizes the evidence of the training data given the Gaussian Process~\cite{gpml}:
\begin{align}\label{marginal_gp_likelihood}
-\log p(\ve{y}|\ve{X},\theta) = \frac{1}{2}\ve{y}^T(\ve{k}(\ve{X},\ve{X})+\sigma_n^2 \ve{I})^{-1}\ve{y} + \frac{1}{2} \log|\ve{k}(\ve{X},\ve{X})+\sigma_n^2 \ve{I}| - \frac{N_s}{2}\log 2\pi~,
\end{align}
where $\ve{I}$ is the identity matrix, and $\sigma_n$ is an arbitrary small level of uncorrelated \emph{noise} included to make the algorithm more numerically stable (possibly in addition to a noise term added into the kernel function to model stochasticity of the original function). Using Bayes' theorem, the product of this likelihood and some prior can then be sampled or (more commonly) simply maximized with respect to the hyperparameters $\theta$.

\section{Surrogate model of the posterior}\label{sec:code}

Our goal is to interpolate an unknown, possibly multi-dimensional log-posterior distribution with a GP, using the mean prediction $\mu(\ve{x})$ of the GP as a best estimate for the distribution's value. Furthermore we want to achieve an accurate estimate for the standard deviation $\sigma(\ve{x})$ in order to compute where to sample next. The nature of the posterior distribution being an (un-normalized) probability distribution implies certain properties/restrictions, that can be incorporated into the GP surrogate model in order to increase the performance of the algorithm and reduce the risk of numerical issues.  These will be discussed in the following.

\subsection{Choice of kernel function}\label{ssec:kernel}

As discussed in section \ref{ssec:gp}, a kernel function with a minimal set of properties will ensure that the GP converges towards the target function (the log-posterior) given a large enough set of samples. However, in order the keep the computational costs low, we aim to use as few samples as possible, and this can be achieved by choosing a kernel function that encapsulates our prior information on the posterior distribution. 

The prior information that we aim to encode is that the log-posterior distribution is deterministic,\footnote{It would be easy to extend this to stochastic functions by adding a noise component to \Cref{anisotropic_rbf}, but posterior density functions of physical data are most commonly deterministic.} and smooth over a characteristic correlation length-scale, that possibly differs between dimensions and is a fraction of the prior size (as we cannot resolve length-scales much larger than the prior). Our default choice in GPry is an anisotropic quadratic RBF kernel multiplied by a constant:
\begin{align}\label{anisotropic_rbf}
    k(\ve{x},\ve{x}')=c^2\cdot\exp\left(-\sum_{i=1}^d\frac{|x_i-x_i'|^2}{2 L_i^2}\right) \ ,
\end{align}
where $c$ is usually called the output-scale, and $L_{i=1,\ldots,N_d}$ are the length-scales.\footnote{If the covariance matrix of the posterior mode that is modelled is approximately known, and that mode is Gaussian enough, one could transform the parameter space using that covariance matrix so as to normalise the Gaussian, in which case the target function is isotropic and we can use a single common length scale, significantly reducing the computational cost of fitting the hyperparameters. In practice, this approach has its own difficulties: even at late stages of learning, the set of training points is too small to compute the covariance matrix via simple Monte Carlo (weighting by their posterior value), and one needs to resort to other approaches, such as fitting a Gaussian to the training, or MC-sampling from the GP (see e.g.\ \cite{paper_1}), the cost of which would likely compensate for the time saved by fitting a single isotropic correlation length-scale.} On top of the choice of the kernel function itself, prior knowledge on the target function is also incorporated in the priors for the hyperparameters. The fundamental assumption is that the length scales should be of an order of magnitude close to that of the posterior modes, and that the latter would be of an order of magnitude not much smaller than that of the prior ranges for the parameters of the posterior. We express this belief as the length-scales being between $0.01$ and $1$ in units of the prior length in each direction. The lower bound ensures that the GP does not overfit during early stages of the learning by fitting each sample individually as a peak on top of the mean of the GP,\footnote{This condition assumes that the size of the mode is larger than about $1/100$th of the prior width in each dimension, which we find reasonably permissive. If this is not the case, either the prior dimensions or the allowed range for the length scales can be re-adjusted.} while the upper bound represents the fact that the size of the prior box should prevent drawing any conclusions on the characteristic length-scale far beyond the region that can be sampled. The prior of the output scale $c$ is chosen to be very broad and allows for values between $0.001$ and $10000$. The $N_d+1$ free hyperparameters $\{c,L_i\}$ are then chosen such that they maximize \Cref{marginal_gp_likelihood}.\footnote{In a full hierarchical Bayesian treatment, instead of maximising we would have to generate a family of GPs with hyperparameters following the likelihood of \Cref{marginal_gp_likelihood}, each of them giving different predictions according to \Cref{mean_gp,cov_gp}. Unfortunately, generating a MCMC sample in order to marginalize over \Cref{marginal_gp_likelihood} as function of $\theta$ is intractable. There have been some attempts at approximate methods \cite{ALoMEuBQ} however even those introduce some computational overhead which we want to avoid. Luckily, as the number of training points of the GP increases we expect \Cref{marginal_gp_likelihood} to get narrower (for sufficiently tame distributions) so that the difference becomes negligible.\label{footnote:MLII}}

\subsection{Parameter space transformations}\label{ssec:transformations}

As a un-normalized probability density, the posterior is a positive function ($p(\ve{x}) \geq 0$ everywhere), and even for a simple one-dimensional Gaussian, it varies over multiple orders of magnitude. Both enforcing positivity and reducing the dynamic range of function values can be achieved by modeling the result of a power-reduction operation $P\left(p(\ve{x})\right)$ on the posterior (e.g.\ a logarithm \cite{ALoMEuBQ} or a square root \cite{Henning_inference}). We use a $\log$-transformation, since in physics it is very common for likelihoods to belong to the \textit{exponential family} of probability distributions \cite{exponential_class} and in practice many likelihood codes usually return log-probabilities.

Another advantage of modelling in log-space, that was pointed out in \cite{ALoMEuBQ}, is that the characteristic length scale of isotropic kernels (e.g. Radial Basis Function (RBF) or Mat\'{e}rn) tends to be larger, which implies that the GP surrogate better generalizes to distant parts of the function, making the GP more predictive.

In practice, we construct a surrogate model for $\log p(\ve{x})$ given some training samples $\ve{y}=\log p(\ve{X})$. In addition, at every iteration of the algorithm, we \emph{internally} re-scale the modeled function using the mean and standard deviation of the current samples set as
\begin{align}
    \log \widetilde{p}(\ve{X}) = \frac{\log p(\ve{X})-\overline{\ve{y}}}{s_{\ve{y}}}\, ,
\end{align}
where $\overline{\ve{y}}$ and $s_{\ve{y}}$ are the sample mean and standard deviation respectively. This re-scaling acts like a non-zero mean function, causing the GP to return to the mean value far away from sampling locations. This in turn encourages exploration when most samples are close to the mode and exploitation when most samples have low posterior values. This effect can be seen in \Cref{fig:1d_acquisition_kb} where the mean of the GP is pushed to higher values close to the edge of the prior. The variance reduction through division by $s_{\ve{y}}$ aids in ensuring numerical stability by restricting the range of values in the training set.

As for the space of parameters $\ve{x}$, we transform the samples such that the prior boundary becomes a unit-length hypercube. For unbounded priors, such as Gaussian or half-Gaussian, we choose the prior boundary such that it contains a large fraction of the prior probability mass ($99.95\%$ by default, which is usually sufficient for the usual few-$\sigma$ \CL{} contours).

This parameter transformation aims at forcing posterior modes to have similar sizes in all dimensions. This usually leads to comparable correlation length scales of the GP across dimensions, which increases the effectiveness of the limited-memory Broyden-Fletcher-Goldfarb-Shanno (L-BFGS-B) constrained optimizer \cite{L-BFGS-B}, used to optimize both the GP hyperparameters and the acquisition function.

Henceforth, if not specified otherwise, in the context of the training set we will refer to $\ve{X}$ (or $\ve{x}$) as the un-transformed values of the sampling locations while $\ve{y}$ refers to the un-transformed values of the log-posterior distribution at $\ve{X}$.

\subsection{Treatment of infinities and extreme values}\label{ssec:svm}

The log-posterior function is bound to return minus infinity for parameter values far away from the region of interest (the posterior modes): the negative log-posterior can be too large to be represented as a floating point number, or the physics code used to compute the likelihood could fail and report a zero-valued likelihood.

This is valuable information but unfortunately we cannot simply add those infinite values to the GP as \Cref{mean_gp,marginal_gp_likelihood} become ill-defined. Hence we are forced to find some numerically stable way of incorporating this information. Naively one could simply swap out the infinities with some large negative value. This approach turns out to be rather problematic as it introduces a discontinuity in the posterior shape or at least one of it derivatives thus modifying the hyperparameters of the GP's kernel. If we instead ignore these points, the learning algorithm will repeatedly try to acquire points in their vicinity, hence getting stuck.

Our solution to this problem is to simultaneously exclude these \emph{infinities} from the GP, and to use them to divide the parameter space into a \emph{finite} and an \emph{infinite} region using a support vector machine (SVM) classifier \cite{Murphy2012, Cortes1995}.\footnote{A similar "safe exploration space" approach, using different tools, has also been used e.g.\ \cite{safeopt, safeopt_2} in the context of Bayesian optimization.} A SVM defines a hyperplane which maximizes the separation between samples with locations $\ve{x}_i$ belonging to one of two classes $y\in\{-1,1\}$. By defining the distance between points through a kernel function $k(\ve{x}, \ve{x}')$ the separating hyperplane is drawn in a higher-dimensional space which is connected to the sample space by a non-linear transformation. This effectively transforms the separating hyperplane into more complex hypersurfaces which are better suited to the classification problem at hand.

\noindent The categorical predictions $\hat{y}(\ve{x})$ of the SVM are then given by
\begin{align}\label{svm_prediction}
    \hat{y}(\ve{x}) = \mathrm{sgn}\left(b+\sum_{i=1}^{N_s} \alpha_i k(\ve{x}_i, \ve{x})\right)
\end{align}
where the hyperparameters $b$ and $\alpha_i$ are optimized in the training procedure.

\begin{figure}[t]
    \centering
    \input{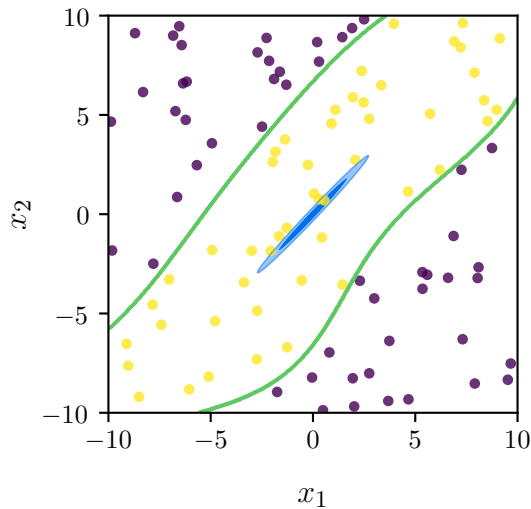}
    \vspace*{-2em}
    \caption{Illustration of the SVM classification. Yellow dots correspond to uniformly sampled locations where the log-posterior distribution is finite while purple dots correspond to infinite log-posterior samples. The green lines are the boundary found by the SVM separating the finite and infinite regions. The blue contours show the 1- and 2-$\sigma$ contours of the posterior distribution (in this case a correlated 2-d Gaussian). In our construction this finite region is designed to roughly correspond to the $10 \sigma$ volume of the Gaussian distribution.}
    \label{svm_illustration}
\end{figure}

We simply use the prediction of the SVM of whether a point is classified as being finite~\mbox{($\hat{y}=+1$)} or infinite~\mbox{($\hat{y}=-1$)} to \enquote{correct} the prediction of the GP. Compared to \Cref{mean_gp,cov_gp} we can explicitly write
\begin{align}\label{mean_gp_svm}
\mu_{\mathrm {GP+SVM}}(\ve{x}) = \mu(\ve{x}) \cdot \begin{cases}
1\ &\mathrm{if}\ \hat{y}(\ve{x}) = +1 \\
-\infty\ &\mathrm{if}\ \hat{y}(\ve{x}) = -1
\end{cases}
\end{align}
For now, we assert such classification from the SVM with absolute certainty, and set
\begin{align}\label{cov_gp_svm}
\Sigma_{\mathrm{GP+SVM}} = \Sigma(\ve{x})\cdot
\begin{cases}
1\ &\mathrm{if}\ \hat{y}(\ve{x}) = +1 \\
0 &\mathrm{if}\ \hat{y}(\ve{x}) = -1
\end{cases} \ .
\end{align}
The precise way of cutting the covariance is irrelevant in our case.\footnote{Still, one could imagine using the SVM output before $\mathrm{sgn}$ function (the classification step) to more smoothly suppress both mean and covariance, possibly combined with a sigmoid function.} \Cref{svm_illustration} shows a two-dimensional toy example of such a classification for a Gaussian distribution in a comparatively much larger prior region.

Aside from making the acquisition procedure more efficient by ignoring unimportant regions, this approach also keeps the overhead cost of the algorithm lower than including a regularized version of the inifinities in the GP. This is because the computational expense of training a SVM scales as $N_s^2$, which is smaller than the $N_s^3$ scaling for the GP.

It is important to recognize that this discussion can also apply not only to infinite values, but also to exceedingly low posterior values. These can be undesirable as they can dramatically change the scale of the emulation problem even though they do not provide a large amount of additional information. In this sense, the algorithm also benefits from a regularization of forwarding too small log-posterior values to the SVM.

We accomplish that by treating all values where $\log p(\ve{x})$ is smaller than some (sufficiently low) threshold as infinities.  However, one has to be careful about the un-normalized nature of the posterior when applying the threshold. In practice, we compare against the maximum of the posterior in the training sample (corresponding to point $\ve{x}_\mathrm{max}$) and only treat values as infinite when $\log p(\ve{x}) < \log p(\ve{x}_\mathrm{max})-T$. We provide a default value for $T$ based on the prescription of \Cref{app:threshold}, also giving the user the option to set it manually.

\section{Learning strategy}\label{sec:strategy}

In \Cref{sec:code} we have described the process of constructing a Gaussian process to emulate the log-posterior distribution once a given set of samples are known. As discussed, a sufficiently large naive set of samples (e.g. prior samples) will in general lead to an accurate model. Unfortunately the computational cost of the algorithm scales with the number of samples $N_s$, both directly as the number of times a possibly-costly true posterior needs to be evaluated, and indirectly by increasing the computational cost of the Gaussian Process itself (as $\sim N_s^2$ at evaluation, and $\sim N_s^3$ when fitting). In practice, samples are chosen so that their location maximises an \emph{acquisition function}, representing some measure of how valuable they would be for the emulation when added to the GP. We discuss this approach in \Cref{ssec:acquisition}. A further reduction in computational costs can be achieved by taking advantage of the number of machines/CPUs in computing clusters (and of CPU cores in user-level CPUs). Thus, we discuss the parallelization of the algorithm in \Cref{sec:paralell}. Finally, in \Cref{ssec:convergence} we discuss the vital question of when to end the acquisition of further samples automatically. Together, this allows \texttt{GPry} to tackle the emulation of arbitrary distributions in a highly parallelized way without relying on the end-user to optimize the number or locations of the samples.

\subsection{Acquisition function}\label{ssec:acquisition}

As discussed above, in order to find a small, but informative set of sampling locations, we will look for locations that maximize an \textit{acquisition function} $a(\ve{x})$. This function will be constructed using a combination of the mean and variance of the GP estimate, in such a way that it balances exploration of the full parameter space (typically where the uncertainty in the prediction is high) with exploitation of areas of high posterior values (which should be more precisely modeled).

\subsubsection{Choice of the acquisition function}

A simple ansatz for an acquisition function that balances exploration and exploitation could be the product of the estimated posterior $p(\ve{x})$ (which is always positive) and its uncertainty $\sigma_{p}(\ve{x})$:
\begin{align}
a_p(\ve{x}) = p(\ve{x}) \cdot \sigma_{p}(\ve{x})\,.
\end{align}

Given that the GP models $\log p$, we have to convert the GP's mean $\mu(\ve{x})$ and uncertainty $\sigma(\ve{x})$ into those of the linear $p(\ve{x})$. Since the transformation from $\log p$ to $p$ is non-linear, the corresponding prediction for $p$ from the GP is not a Gaussian distribution and the computation of its mean and standard deviation is non-trivial. However, in practice these details are irrelevant since the acquisition function only needs to approximate the most beneficial sampling location. Then, we can simply write for the mean $p(\ve x) \approx \exp[\mu(\ve{x})]$ and for the uncertainty $\sigma_p \approx \exp[\mu(\ve{x}) + \sigma(\ve{x})] - \exp[\mu(\ve{x})]$\,. With this, the acquisition function above becomes:
\begin{equation}\label{intermediate_acquisition_function}
  a_p(\ve{x}) \approx
  \exp[2\mu(\ve{x})] \left\{\exp[\sigma(\ve{x})]-1\right\}
  \,,
\end{equation}
which is similar to the acquisition functions used in \cite{acquisition_functions_1,acquisition_functions_2}. This approximation can be further linearized assuming $\sigma(\ve{x}) \ll 1$ to give $a^\mathrm{lin}_p(\ve x) = \exp[2\mu(\ve x)] \sigma(\ve x)$. 

As discussed in the next section, we found it beneficial to boost the exploratory behaviour of the acquisition function, especially in high dimensions. To achieve that, we include a relaxation factor $\zeta \in (0,1]$ multiplying the mean to discourage exploitation \final{(similar to what was done in e.g.\ \cite{paper_1})}. This yields the final acquisition function:
\begin{equation}\label{final_acquisition_function}
  a_p(\ve{x}) \approx
  \exp[2\zeta\mu(\ve{x})] \left\{\exp[\sigma(\ve{x}) - \sigma_n]-1\right\}
\end{equation}
which can again be linearized as $a^\mathrm{lin}_p(\ve{x}) = \exp[2\zeta\mu(\ve{x})] [\sigma(\ve{x}) - \sigma_n]\,.$ Notice also that from the $\sigma(\ve{x})$ term we have subtracted a possible uncorrelated noise term proportional to $\sigma_n^2$ in the kernel function (equivalently, a constant term added to the diagonal of the kernel covariance matrix). This is because for acquisition purposes we only care about the uncertainty coming from decorrelation from the sampled locations.

The logarithm of this acquisition function is maximized at every acquisition step which yields a candidate for the next sampling location. The computational overhead of the acquisition procedure is dominated by the prediction of $\mu(\ve{x})$ and $\sigma(\ve{x})$ by the GP which scales as $\sim N_s^2$.

\subsubsection{Acquisition hyperparameter}
\label{ssec:zeta}

The effect of $\zeta$ in \Cref{final_acquisition_function} is that of balancing exploitation and exploration. Values of $\zeta$ that are too high make the algorithm focus too much on the top of a posterior mode, so that samples in the tails are unlikely to be proposed, and during large numbers of iterations the GP model is mostly stable (only adding high-posterior but low-information samples). This leads to unnecessarily high computational costs, and often to false positives in assessing convergence. These effects are more dramatic in higher dimensions. On the other hand, values of $\zeta$ that are too low would produce more regular but slower convergence, neglecting information about the expected value of the true function that could have been exploited to converge faster. In general, a sub-optimal choice of $\zeta$ will increase the amount of samples necessary for convergence, sometimes quite significantly.

\begin{figure}[t!]
    \centering
    \includegraphics[width=0.9\textwidth]{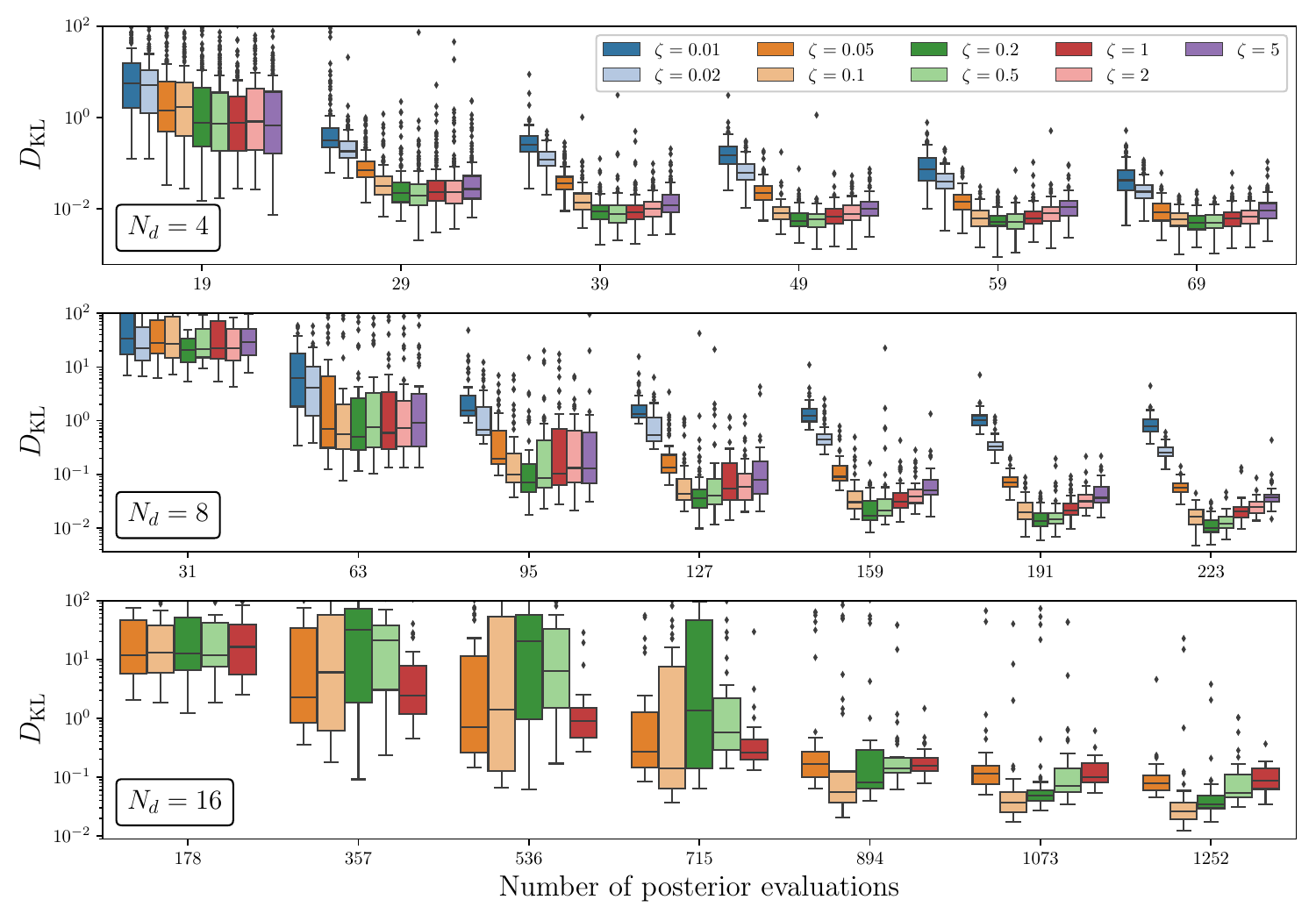}
    \caption{
    Distribution of Kullback-Leibler divergences between the GP prediction and the true distribution at various learning stages (i.e., $N_s$ samples) for random correlated Gaussian posteriors with dimensionality $N_d=4, 8$, and $16$ (150, 50, and 50 realisations, respectively). The boxes represent inter-quartilic ranges, the black line inside them the median, and the whiskers and dots represent the tails of the distributions. For each dimensionality there is a visible trend towards an optimal trade-off between exploration and exploitation in terms of $\zeta$.}
    \label{fig:zeta_vs_kl}
  \end{figure}

To select appropriate values, we have conducted a series of experiments on degenerate Gaussian posterior distributions in 2, 4, 8 and 16 dimensions (for $N_d<4$ the effect of $\zeta$ is small), generated as explained in \Cref{ssec:multivariate_gaussians}. In order to isolate the effects of $\zeta$, in these tests we have not used the parallelization scheme described in \Cref{sec:paralell}. The results of these experiments in terms of KL divergence (see \Cref{app:kl}) are shown in \Cref{fig:zeta_vs_kl}, and have led us to propose the empirical formula $\zeta = N_d^{-0.85}$ as a default value for $\zeta$ (users can override it if prior knowledge of the posterior shape suggests that exploration should be favored over exploration or vice versa). Preliminary tests in higher dimensions (up to $N_d=27$) have shown this formula to produce good results. The fact that a fixed $\zeta$ becomes greedier as dimensionality goes up should not come as a surprise, as discussed in \Cref{app:threshold}.

\subsubsection{Optimization of the acquisition function}\label{sec:opt_of_acq}

For the maximization of the acquisition function we use the L-BFGS-B optimizer \cite{L-BFGS-B} included in the \texttt{scipy} Python package. Since this optimization problem is highly non-convex, with the acquisition function often having many disconnected maxima, the numerical optimization is performed multiple times from different randomly-drawn starting locations. In high dimensions drawing an initial point with a non-vanishing value of the acquisition function becomes increasingly unlikely as the prior volume with vanishing posterior increases as a power of the dimension (curse of dimensionality).

Because of this problem, optimal proposals most likely fall in the vicinity of the current sampling locations. In order to generate such points we, by default, use a \emph{centroid} algorithm: take the average location of $N_d+1$ randomly selected samples, and perturb them in each dimension by the coordinate difference to one the samples multiplied by a draw from an exponential distribution with parameter $1/\lambda$. Here a lower $\lambda$ increases the spread of the proposed locations. A fraction of the locations are drawn from a uniform distribution within the original prior boundaries, in case a region of high posterior has not yet been captured by the current samples.

For highly non-Gaussian distributions this method of proposing points tends not to be exploratory enough. In these cases we resort to drawing proposals uniformly within the prior volume.

Lastly also provide a method to generate Gaussian-distributed proposals given an estimate of the mean and covariance matrix of the posterior, if such information happens to be known.

\final{
We notice that alternative approaches to maximizing the acquisition function exist. In \cite{paper_1}, the are picked out of an MCMC of the mean GP model.}

\subsection{Parallelization}\label{sec:paralell}

The naive approach of using the acquisition function presented above is to acquire and evaluate sampling locations in sequence, with each acquisition step consisting of the evaluation of the true posterior distribution (and updating the GP model) in order to obtain the next candidate for a sampling location. However, as often multiple processing units (either on the same or across different machines) are available, we can make this algorithm more efficient by attempting to propose \emph{batches} of sampling locations, so that the true posterior, which is expected to be the largest source of computational cost, can be evaluated in parallel.

There have been many different proposals for batch acquisition for GPs in the past which can broadly divided into two categories:

Algorithms like \cite{parallel_1, parallel_2, parallel_kriging_believer_1, parallel_kriging_believer_2} construct an acquisition function which can be optimized for several points at once. However, for a $d$-dimensional posterior distribution acquiring $q$ points at once involves global optimization in $d\cdot q$ dimensions which obviously becomes computationally prohibitive even if $d$ and $q$ are not extremely large.

The second category \cite{parallel_3, parallel_kriging_believer_1, parallel_kriging_believer_2} works by sequentially acquiring multiple points without having to sample from the posterior distribution in between and afterwards evaluating the true posterior at the gathered locations in parallel. We will be using one of these methods called the \textit{Kriging believer} method \cite{parallel_kriging_believer_2}. 

% journal will remove spaceing! let's try without
%\vspace*{-0.5\baselineskip}
\subsubsection*{The Kriging believer method}
%\vspace*{-0.5\baselineskip}

\begin{figure}[h]
\begin{center}
\includegraphics[width=\textwidth]{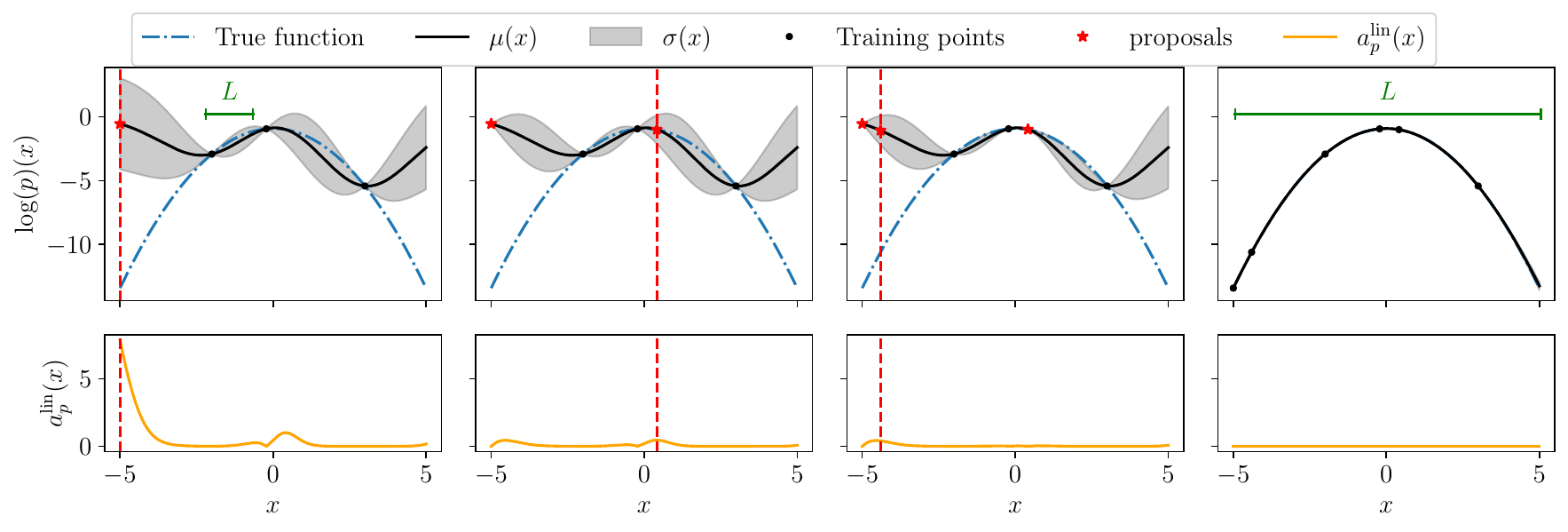}
\end{center}
\caption{Illustration of the Kriging believer method. Three points are acquired sequentially (three left plots) by using the prediction from the GP instead of evaluating the posterior at each iteration. After the three samples have been acquired the posterior function can be evaluated at these points (right). The hyperparameters of the GP regressor only need to be refit at the last step.
Obviously using this approach comes at the expense of requiring more points to converge (e.g.\ the third point did not add much information and is unlikely to have been selected after the second one if using sequential acquisition). This can however be compensated by the computation time that is saved by both acquiring points faster and evaluating the posterior in parallel. The characteristic length-scale $L$ of the kernel increases as more samples are added, which aids the better fit in the right panel.}
    \label{fig:1d_acquisition_kb}
\end{figure}

The fundamental assumption of the Kriging believer method (similarly to our assumption when constructing the acquisition function) is that the value of the posterior distribution in any point roughly equals the predicted mean of the GP. We can therefore acquire a batch of points by sequentially (1) obtaining a maximum of the acquisition function at $\ve{x}_*$, (2) assuming for it a log-posterior evaluation equal to $\mu(\ve{x}_*)$, (3) adding it to an intermediate \emph{augmented} GP (thereby producing a different new maximum of the \emph{augmented} acquisition function), and repeating until the desired number of locations has been proposed. This method will be increasingly accurate as more samples are added to the GP so that $\mu(\ve{x}_*)$ approaches the true $\log p(\ve{x}_*)$. An illustration of the Kriging believer algorithm sampling on the log of a normal distribution is shown in \Cref{fig:1d_acquisition_kb}. 

The obvious advantage of this method, as discussed above, is that the true posterior can be evaluated in parallel for the acquired locations. This is beneficial as we expect the true posterior evaluations to dominate the computational cost in most scenarios. In addition, there is another source of speedup: since adding new mean-valued samples does not change the optimal hyperparameters of the GP according to \Cref{marginal_gp_likelihood}, there is no point to re-fitting them (see \Cref{ssec:timecosts}).\footnote{On top of that, the necessary step of inverting the kernel matrix after adding new points, in order to get predictions using \Cref{mean_gp,cov_gp}, could be accelerated by taking advantage of the fact that the inverse of the previous kernel matrix is known, using a fast, blockwise matrix inversion formula. To our knowledge this has not been pointed out in the past. In our case, the amount of possible time savings is small.}

In terms of the precise size of the batch, there is evidently a trade-off between the speedup gained by not refitting the GP's hyperparameters at every iteration, and the inaccuracy of the GP's mean prediction making the active sampling less efficient as the number of Kriging believer steps grow. Due to the loss of accuracy in the predictions, more samples are required to converge to the true distribution, but this is compensated by the speedup achieved through parallel evaluations of the posterior. Overall this results in a smaller number of iterations (hence a smaller wall-clock run time) than sequential learning, as long as the size of the batches is kept reasonable. 

We find that a batch size corresponding to at most the number of dimensions of the inference problem $N_d$ works reasonably well. We therefore set the standard number of Kriging believer steps to the minimum between $N_d$ and the number of parallel processes.

\subsection{Convergence criterion}\label{ssec:convergence}

The last component of our algorithm is its \emph{convergence criterion}, which should terminate it as soon as (or at least not much later than) the GP has reached sufficient precision at modelling the log-posterior. Precision could be assessed as the reduction in the variance of a GP-predicted global posterior quantity such as the evidence $\int p(\ve{x})\,\mathrm{d}\ve{x}$. Analytical computation of these quantities in terms of the GP are usually not possible, e.g.\ in our case because of the modelling of the log-posterior instead of the posterior itself, or because the product of a GP times arbitrary priors does not have a closed-form integral in general. Numerical approaches would involve MC samples of the GP-modelled posterior, which come at a reasonably-small computational cost, but whose use for the convergence criterion would involve obtaining them at (nearly) every iteration.

A much cheaper convergence criterion would involve computations using the much smaller set of current and/or proposed GP samples. We propose one such criterion, that we call \texttt{CorrectCounter}, based on observing the accuracy of the learning process and stopping when the model does not seem to learn any new information. We will show how that the speedup in this case does not necessarily come at the cost of precision.

The assumption here is that our algorithm stops learning if the the GP's predictions at newly acquired sampling locations $\ve{x}$ repeatedly match the value of the true log-posterior $\log p(\ve{x})$ distribution to close approximation. We set a threshold using relative and absolute tolerances $\epsilon_{\mathrm{abs}}, \epsilon_{\mathrm{rel}}$ such that
\begin{align}\label{eq:CorrectCounter}
    \left|\mu_{\mathrm{GP+SVM}}(\ve{x})-\log p(\ve{x})\right| \overset{!}{<} \epsilon_{\mathrm{abs}} +\left|y_{\max} - \mu_{\mathrm{GP+SVM}}(\ve{x})\right| \cdot\epsilon_{\mathrm{rel}}\,,
\end{align}
where $y_{\max}$ is the largest log-posterior from the current GP sample, and $\mu_{\mathrm{GP+SVM}}(\ve{x})$ is the GP's prediction at $\ve{x}$ before the GP has been fit to this point. This criterion can be computed at virtually no cost, since both $\mu_{\mathrm{GP+SVM}}(\ve{x})$ and $\log p(\ve{x})$ have been computed as part of the acquisition procedure. If this condition is satisfied a few times in a row we consider the model converged and stop the algorithm.  Convergence in this case means a guarantee that (on average) new evaluations of the GP will at least approximately comply with the true posterior at the same location (as opposed to convergence meaning stability of some global quantity).

Similarly to the discussion in \Cref{ssec:svm,ssec:zeta}, the behaviour of this convergence criterion is sensitive to the dimensionality $N_d$ of the problem. As explained in \Cref{app:threshold}, since the dynamic range of a log-posterior enclosing a given probability mass grows with dimensionality, the effect of a constant $\epsilon_\mathrm{abs}$ will become more stringent as dimensionality increases, making the criterion fail to report as converged GP models that already very precisely characterise the posterior. In \Cref{app:threshold} we propose a way to relax $\epsilon_\mathrm{abs}$ in a dimensionally-consistent way. The relative thresold $\epsilon_\mathrm{rel}$ should not be affected by dimensionality, and it is fixed to $0.01$. In both cases, we also give the user the option to set their own values for the convergence criterion.

On the other hand, as the number of dimensions $N_d$ increases, correctly mapping the tails of the distribution becomes increasingly more important (for a detailed discussion see \Cref{app:threshold}), while the surrogate model tends to converge first around the maximum of true posterior distribution. The tails usually remain underrepresented at first and only get explored later in the acquisition procedure. A higher dimensionality therefore makes it likelier to acquire a batch of consecutive correctly-predicted points in a non-converged GP model around the top of the mode. We account for this by increasing the number of times points have to be predicted correctly to claim convergence to $n = N_d/2$ (with the exception of fixing $n = 4$ for low dimensionality, $N_d < 8$). This reduces the risk of neglecting convergence at the tails.

\begin{figure}[h]
\begin{center}
\includegraphics[width=0.49\textwidth]{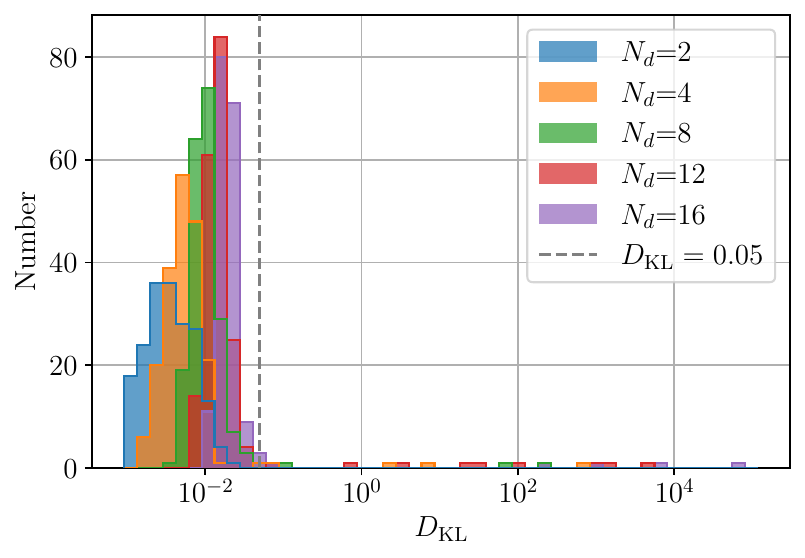}
\includegraphics[width=0.49\textwidth]{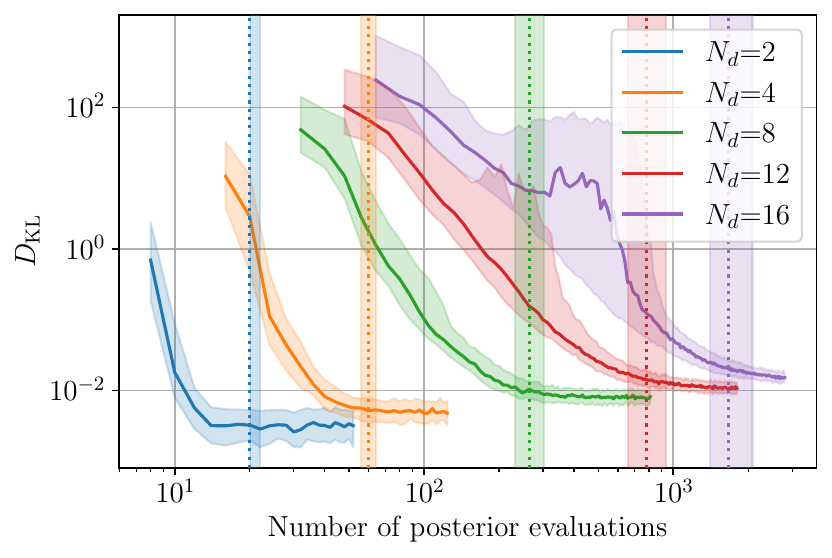}
\end{center}
\caption{\textbf{Left}: Distribution of KL divergences between GP models and their correspoding true posterior at \texttt{CorrectCounter}-reported convergence, for $N_d=2, 4, 8, 12, 16$-dimensional random correlated Gaussians (200 draws per dimensionality). Only a small fraction ($<$5\%) surpass our target value of $D_{\mathrm{KL}}=0.05$.
\textbf{Right}: Medians (solid lines) and interquartile ranges (shaded bands) of the KL divergences between GP models and their correspoding true posterior, for the same sets of Gaussians, as function of their number of accepted (finite) samples. The dashed vertical lines indicate the median number of accepted steps at which \texttt{CorrectCounter} reports convergence, and the shaded vertical bands the respective interquartilic ranges. As expected there is a trend towards higher values of $d_{\mathrm{KL}}$ visible as $N_d$ increases, but it is well under control for the dimensionalities targeted in this study.}\label{fig:correct_counter}
\end{figure}

We tested the \texttt{CorrectCounter} criterion on a set of correlated Gaussians in 2, 4, 8, 12 and 16 dimensions, generated as explained in \Cref{ssec:multivariate_gaussians}. We target a KL divergence with respect to the true Gaussian distribution of less than $5\%$. As shown in \Cref{fig:correct_counter}, we achieve such threshold with the settings described above for the tolerances and the number of consecutive correct predictions, at least for the range of dimensionality targeted in this study. Towards higher dimensionality there is a trend to converge before the convergence curve flattens out entirely, which hints at the need for more sophistication in dealing with dimensional consistency. We leave this for future work.

We also note that we have written a criterion based on the costlier KL divergence (see \Cref{app:kl}), which we provide as an alternative option. This alternative criterion is based on the posterior emulation stabilizing over multiple subsequent steps (defined through the KL divergence being below some critical threshold). This criterion comes with its own sets of challenges, such as incorrectly detecting convergence when non-informative points are added to the GP or the costly nature of its computation. Nonetheless it can be preferable when the log-posterior function is extremely expensive to evaluate or when the posterior distribution exhibits unusual features, as this convergence criterion is not only sensitive to the acquired samples but also to the hyperparameters of the GP.

\section{The full algorithm} \label{sec:full_algorithm}

In this section we present the full structure of the algorithm, entailing the generation of the initial set of training samples (\Cref{ssec:initial_training}), the main acquisition loop that sequentially looks for optimal samples and checks convergence (\Cref{ssec:algorithm}), and the final generation of a Monte Carlo sample of the trained GP surrogate model of the posterior, which can be used to get marginalised quantities (\Cref{ssec:timecosts}), together with a comparison of computational costs of this algorithm against those of classic Monte Carlo.

\subsection{Initial training set}\label{ssec:initial_training}

In order to start the sequential acquisition of points we need an initial training set containing samples from our posterior distribution. These do not have to be very informative samples but need to be finite (according to the definition in \Cref{ssec:svm}) and uncorrelated, in order to generate some very crude but meaningful initial interpolation of the log-posterior distribution.

Of course we want to choose this sample such that the ratio of finite to infinite log-posteriors is reasonably high, in order not to waste too many posterior evaluations on the initial point generation. In low dimensions and with small priors compared to the size of the mode, any random generator (such as draws from the prior itself, or from a uniform distribution within the prior bounds) would produce initial samples satisfying the requirement above. As the ratio of the prior to posterior volume grows with the number of dimensions, randomly drawing a finite point from the prior becomes increasingly unlikely. In this case, prior knowledge of the posterior can be incorporated, usually in the form of a \enquote{reference} distribution which is a rough guess of where the mode might be (the same that is commonly used to generate initial points for MCMC). In general any guess for reasonable parameter values that lead to a finite posterior can be used, which can be obtained from physical considerations of the underlying model.

\subsection{Main algorithm}\label{ssec:algorithm}

In Algorithm~\ref{main_algoirthm} we show the main algorithm used within the \texttt{GPry} tool in pseudo-code, consisting mostly of the optimization and acquisition loops (the latter based on the Kriging believer approach). This pseudo-code mostly summarizes the ideas which are explained in the corresponding \Cref{sec:strategy,sec:code}. 

Note that the $n_{r,\mathrm{GP}}$ starting locations for the optimization of the hyperparameters in Line \ref{alg:gp_fit} are sampled logarithmically in the hypervolume. The step of Line \ref{alg:gp_fit} is currently the most expensive step, scaling as $N_s^3$ due to the required repeated matrix inversion required for computing $\log p(\theta|\ve{X},\ve{y})$. This is why we only perform this step every $n_\mathrm{opt}$-th time, and otherwise we optimize the hyperparameters starting only from the previous best fit. The next most expensive step is the acquisition function optimization in line~\ref{alg:acq_opt}, and scales approximately as $N_s^2$ due to the repeated evaluation of the acquisition function requiring the evaluation of $a(\ve{x})$, which itself requires matrix multiplications.

\IncMargin{1em}
\begin{algorithm}[H]

 \DontPrintSemicolon
 \SetKwFor{RepTimes}{repeat}{times}{end}
 \SetKwIF{EveryNth}{}{EveryNthElse}{every}{time}{}{otherwise}{end}%
 \KwIn{$\ve{X}$ (initial samples), $\ve{y}$ (initial log-posterior values)}
\For{$n<N_\mathrm{max}$}{
  fit SVM with $\ve{X}, \ve{y}$\;
  \uEveryNth{$n_\mathrm{opt}$-$\mathrm{th}$}{
  find $\theta_{\mathrm{MAP}}=\mathrm{argmax}[\log p(\theta|\ve{X}, \ve{y})]$ from $n_{r,\mathrm{GP}}$ starting locations \label{alg:gp_fit} \comment*[r]{\Cref{marginal_gp_likelihood}}
  }
  \EveryNthElse{
  find $\theta_{\mathrm{MAP}}=\mathrm{argmax}[\log p(\theta|\ve{X}, \ve{y})]$ from last best-fit \comment*[r]{\Cref{marginal_gp_likelihood}}
  }
  \texttt{GP\_fit}($\ve{X},\ve{y}$)\;
  $\ve{X}_\mathrm{new} = [\,]$\;
  $\ve{X}_\mathrm{lie}$ = $\ve{X}$ \qquad and \qquad $\ve{y}_\mathrm{lie}$ = $\ve{y}$\;
  \RepTimes{$M$}{
  find $\ve{x}_\mathrm{add}=\mathrm{argmax}[a(\ve{x})]$ starting from $n_{r,\mathrm{acq}}$ starting locations\label{alg:acq_opt}\;
  $\ve{X}_\mathrm{lie}$ \textit{append} $\ve{x}_\mathrm{add}$ \qquad and \qquad $\ve{X}_\mathrm{new}$ \textit{append} $\ve{x}_\mathrm{add}$ \;
  $\ve{y}_\mathrm{lie}$ \textit{append} $\ve{\mu}(\ve{x}_\mathrm{add})$ \comment*[r]{Kriging believer}
  \texttt{GP\_fit}($\ve{X}_\mathrm{lie}, \ve{y}_\mathrm{lie}$)\;
  }
  $\ve{y}_\mathrm{true} = \log \mathcal{L}(\ve{X}_{\mathrm{new}})+\log\pi(\ve{X}_{\mathrm{new}})$\comment*[r]{parallelizable}
  $\ve{X}$ \textit{append} $\ve{X}_\mathrm{new}$\;
  $\ve{y}$ \textit{append} $\ve{y}_\mathrm{true}$\;
  \lIf{\texttt{is\_converged} $\mathrm{(e.g.~\Cref{eq:CorrectCounter})}$}{\textbf{break}}
 }
 Sample $\mu(\ve{x})$ with MC sampler\;
 \KwRet{MC sample}
\nonl{\;}
\nonl{\;}\;
\SetKwProg{Fn}{Function}{}{end}
\Fn{\texttt{GP\_fit($\ve{X},\ve{y}$)}}{
Compute $K^{-1}=\ve{k}(\ve{X},\ve{X}|\theta_{\mathrm{MAP}})^{-1}$\comment*[r]{matrix inversion}
$\mu(\ve{x})=\mu_{\mathrm{GP+SVM}}(\ve{x})$\comment*[r]{ \Cref{mean_gp,mean_gp_svm}}
$\sigma(\ve{x})=\sqrt{\Sigma_{\mathrm{GP+SVM}}(\ve{x})}$\comment*[r]{\Cref{cov_gp,cov_gp_svm}}
$a(\ve{x}) = \exp[2\zeta \mu(\ve{x})] \{\exp[\sigma(\ve{x})-\sigma_n]-1\}$ \comment*[r]{\Cref{final_acquisition_function}}
}

\caption{
The \texttt{GPry} algorithm in a condensed format, omitting the internal transformations that are made to the data. $M$ is the number of Kriging believer steps made in each iteration. The overhead of the algorithm is dominated by the computations performed in Lines \ref{alg:acq_opt} and \ref{alg:gp_fit}.
}\label{main_algoirthm}
 \end{algorithm}

\subsection{Modelling the marginalized posterior}\label{ssec:timecosts}

As mentioned above, to compute marginalized 1D/2D posteriors, we have to compute a high-dimensional integral of our emulated posterior (see \Cref{ssec:bayesian_inference}). This can be achieved by integrating the GP numerically through the creation a Monte Carlo sample, either based on nested sampling, Metropolis Hastings sampling, or (using the backward differentiable nature of the GP) even Hamiltonian sampling. As \texttt{GPry} is interfaced with the \texttt{Cobaya} package \cite{cobaya}, its standard samplers can also be used to generate the final MCMC sample. Currently, this sampling is performed using the GP's mean prediction according to \Cref{mean_gp_svm} as the posterior distribution to sample.\footnote{Technically, the information that is available through the covariance of the GP could be used to obtain an estimate of the uncertainty of emulation on our final posterior sample. As the acquisition procedure only stops if the posterior mode is mapped accurately enough, this assures that at convergence this variance is sufficiently small to safely be neglected.}

One important question that such an approach poses, however, is whether the emulation of the posterior with the GP with subsequent sampling of the surrogate posterior will be computationally more efficient than the direct sampling of the true posterior. For this, let us use a simple back-of-the-envelope computation. Consider the time to run a full sampling of the true posterior as $N_t t_t$\,, where $t_t$ is the approximate time for a single evaluation and $N_t$ the total number of samples required. Instead, the time to run a sampling of the GP posterior can be estimated as $N_g t_g$\,, where $t_g$ is the average time for a single GP evaluation and $N_g$ the total number of required GP samples. Additionally, and crucially, there is the additional overhead of constructing the GP in the first place, which we will denote simply as $T_o$ for now (we will discuss this in more detail later). In that case, the construction of a GP is advantageous if
\begin{equation}
    T_o + N_g t_g < N_t t_t
\end{equation}
Typically it can be assumed that $t_g \ll t_t$ except for very simple toy models. Furthermore, typically $N_g \simeq N_t$ if one uses MCMC/nested sampling methods to sample the GP, or even $N_g \ll N_t$ if one can use Hamiltonian MC methods on the GP but not on the true posterior. Thus, as long as $T_o$ remains reasonably lower than $N_t t_t$ (the total runtime of the MCMC), the use of a GP would always be advantageous. It is thus crucial to obtain a precise estimate for the overhead time $T_o$. This overhead depends strongly on the dimensionality of the problem, the non-Gaussianity of the posterior, and the underlying machine executing the code.

Looking at the timing information from the multi-variate Gaussian cases of \Cref{ssec:multivariate_gaussians}, the overhead was dominated by the numerical optimization of the acquisition function (Line~\ref{alg:acq_opt} of Algorithm~\ref{main_algoirthm}), taking very roughly $100s \cdot (N_s/100)^{2.4}$ (to give an order-of-magnitude estimate). The next most important factor, the optimization of hyperparameters (Line~\ref{alg:gp_fit} of Algorithm~\ref{main_algoirthm}) only takes around $3s \cdot (N_s/100)^{3.2}$ (order of magnitude) in total. It has a smaller pre-factor since it is only performed every $n_\mathrm{opt}$-th iteration, while the acquisition optimization is performed $M \cdot n_\mathrm{r,acq}$ times per iteration, see Algorithm \ref{main_algoirthm}. It is thus comparatively irrelevant for $N_s \ll 10^4$, which is almost always the case for the range of dimensionalities considered in this study.

\begin{figure}
    \centering
    \includegraphics[width=0.45\textwidth]{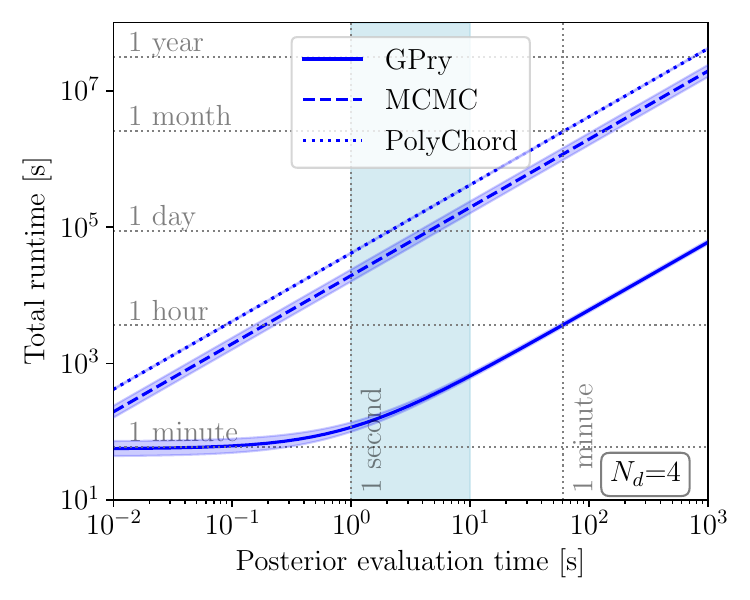}
    \includegraphics[width=0.45\textwidth]{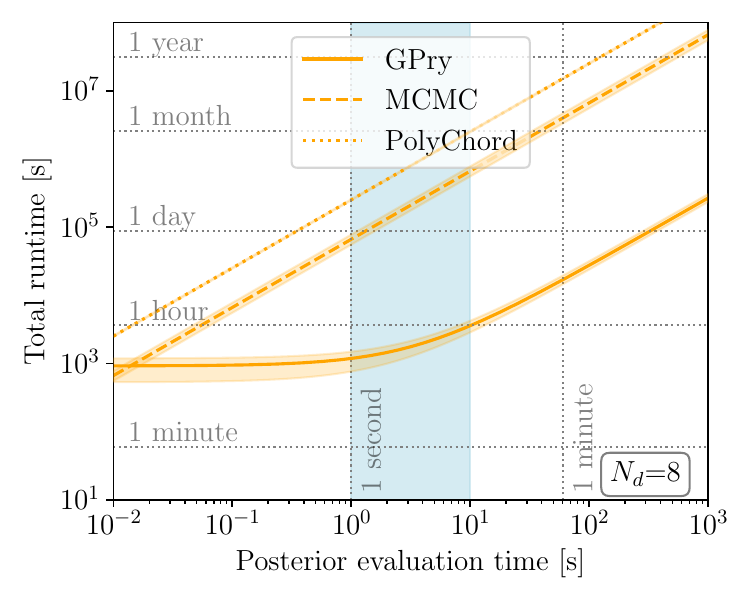}
    \includegraphics[width=0.45\textwidth]{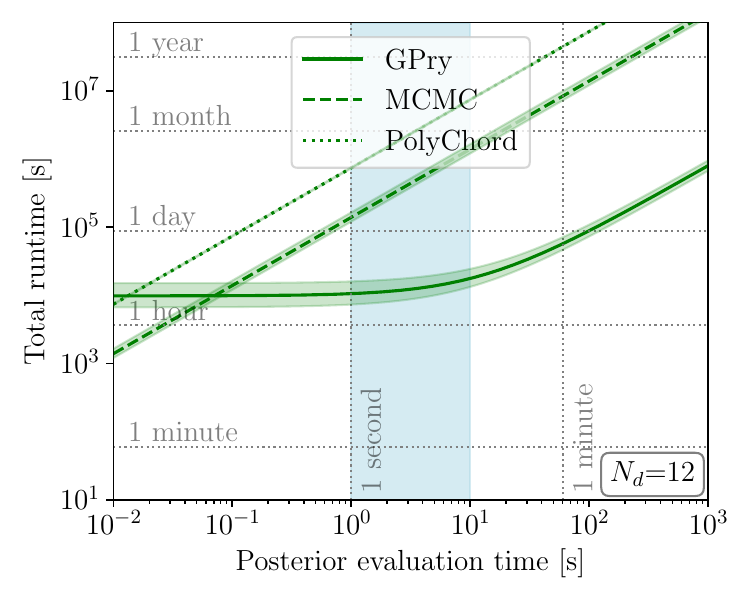}
    \includegraphics[width=0.45\textwidth]{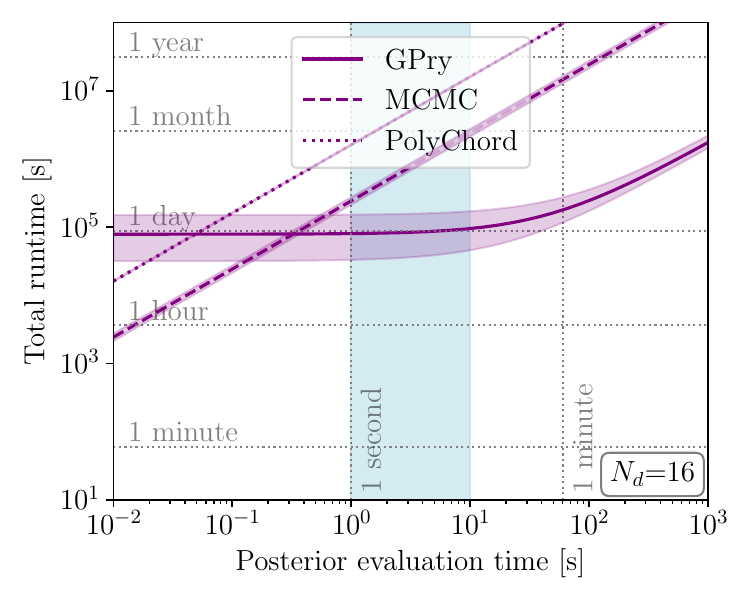}
    \caption{Order of magnitude estimate of total runtime comparison of \texttt{GPry} with the MCMC sampler \texttt{CosmoMC/Cobaya} and the nested sampler \texttt{PolyChord} (via its \texttt{Cobaya} interface). The comparison is done for multi-variate Gaussians of various dimensionalities, and shows the median as a line and the 25\% and 75\% quantiles as a shaded area. The comparison is run with only a single CPU, but the orders of magnitude hold for similar computational resources for all three methods. The light blue band gives an approximate range of computation times of standard cosmological codes (like \texttt{camb} or \texttt{CLASS}) which depend strongly on the considered model and observables. Note that while MCMC and \texttt{PolyChord} are dominated by the posterior evaluation time everywhere, \texttt{GPry} is dominated by overhead for small posterior evaluation times.}
    \label{fig:timings}
\end{figure}

In \Cref{fig:timings} we report the approximate expected total runtime of \texttt{GPry} compared to the \texttt{Cobaya} implementation of the MCMC sampler \texttt{CosmoMC} \cite{mcmc_sampler_1, mcmc_sampler_2, cobaya} and the nested sampler \texttt{PolyChord} \cite{polychord_1, polychord_2} (via its \texttt{Cobaya} interface), neglecting the overhead of MCMC and \texttt{PolyChord} as in this case it is tiny \cite{cobaya}. For example, in the case of a $N_d=12$ multi-variate Gaussian, \texttt{GPry} would outperform the MCMC (which requires $\approx1.5\cdot 10^5$ evaluations) for posterior evaluation times larger than $\sim 0.1$ seconds. Comparing to the average runtime of a cosmological code such as \texttt{CLASS}, on average we find a significant speedup all the way up to 16 dimensions.

Note that in \Cref{fig:timings} we show single-core performance with as many Kriging believer steps as dimensions (while still evaluating the posterior sequentially). The curves shown for MCMC and especially for \texttt{PolyChord} would drop almost proportionally to the number of cores available, while \texttt{GPry} does not scale quite as well. However, for a similar amount of computational resources, up to a number of processes similar to the dimensionality of the problem, these results are expected to hold in order of magnitude. While the runtime of MCMC and \texttt{PolyChord} is dominated by the posterior evaluations, the overhead of \texttt{GPry} is considerable and might scale differently depending on the underlying architecture. Further improvements in runtime could be made by optimizing the underlying GP implementation.

%%%%%%%%%%%%%%%%%%%%%%%%%%%%%%%%%%%%%%%%%%%%%%%%%%%%%%%%%%%%%%%%%%%%%%%%%%%%%%%%
\section{Examples}\label{sec:examples}

After having discussed the design of the \texttt{GPry} code in \Cref{sec:code,sec:strategy,sec:full_algorithm}, we now demonstrate the performance of the code using a variety of examples, both Gaussian and non-Gaussian distributions considered in the literature, as well as examples from cosmological applications. 

\subsection{Multi-variate Gaussians}\label{ssec:multivariate_gaussians}
The example of a multi-variate Gaussian distribution is enlightening as a benchmark for the average performance of the GP, as it can quite trivially be scaled with dimensionality and many likelihood functions can - at least around their maximum - be reasonably well approximated by Gaussian distributions. We can thus use it as a benchmark for performance and accuracy as a function of dimensionality, as well as to model critical scalings such as that of the $\zeta$ parameter from \Cref{ssec:acquisition}, the factors involved in \Cref{eq:CorrectCounter}, and the timings relevant for \Cref{ssec:timecosts} (see \Cref{app:threshold}).

\noindent We generate correlated multidimensional Gaussians with log-likelihood function
\begin{align}\label{eq:randomly_correlated_gaussians}
    \log \mathcal{L}(x_0, \dots, x_n) = -\frac{(\ve{x}-\mathfrak{m})^T\ve{C}^{-1}(\ve{x}-\mathfrak{m})+\log((2\pi)^n|\ve{C}|)}{2}
\end{align}
by drawing a random covariance matrix that satisfies
\begin{align}
    \ve{C}_{i,j} = \sigma_i \sigma_j \mathrm{corr}_{i,j}
\end{align}
where $\mathrm{corr}_{i,j}$ is a randomly drawn correlation matrix with uniformly drawn eigenvalues,\footnote{They are uniformly drawn between 0 and 1, then multiplied by a normalization constant such that their sum equals the number of dimensions, in order to avoid cases where many of the eigenvalues are close to zero simultaneously.} and the standard deviations are uniformly drawn as  $\sigma_i\in[0,1]$. The mean vector $\mathfrak{m}$ is set to 0 and the prior fixed to $5\sigma_i$ in each direction. This ensures that the mode is centered within the prior. The case in which parts of the mode are cut off by the prior is discussed in section \ref{sec:non-gaussian_distributions}. We then conducted tests in $\{2, 4, 8, 12, 16\}$ dimensions recording the the Gaussian KL divergence of \Cref{eq:kl_gaussian}, the number of posterior evaluations, and the overall overhead. The final results were already shown in \Cref{fig:correct_counter,fig:zeta_vs_kl,fig:timings}.

\subsection{Non-Gaussian distributions}\label{sec:non-gaussian_distributions}
One of the main goals of our algorithm is to be robust with regards to the functional shape of the posterior distribution. We therefore tested the code also on non-Gaussian distributions with varying degrees of pathological features. All adopted priors are flat in the respective parameters.

\subsubsection{Log-transformations}
Our first example of a non-Gaussian feature is motivated by a common occurrence in Physics. In many applications, there are free scales in the problem which are not known across one or more dimensions in the parameter space. For these parameter one usually samples their logarithm with a flat prior (which is equivalent to imposing a logarithmic prior), distributing the prior probability density evenly across multiple orders of magnitude. If the likelihood is Gaussian in the (linear) parameter, this typically leads to a log-Gaussian distribution of the form
\begin{align}
    10^x \sim \mathcal{N}(\mu,\sigma)
\end{align}
across some dimensions.

\begin{figure}
    \centering
    \includegraphics[width=0.45\textwidth]{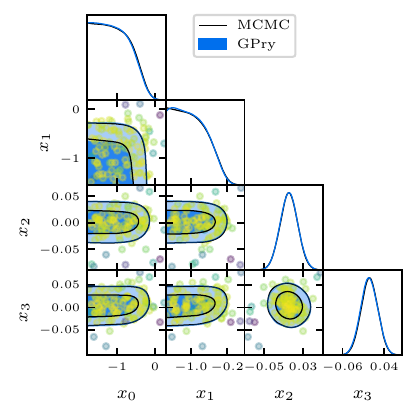}
    \includegraphics[width=0.54\textwidth]{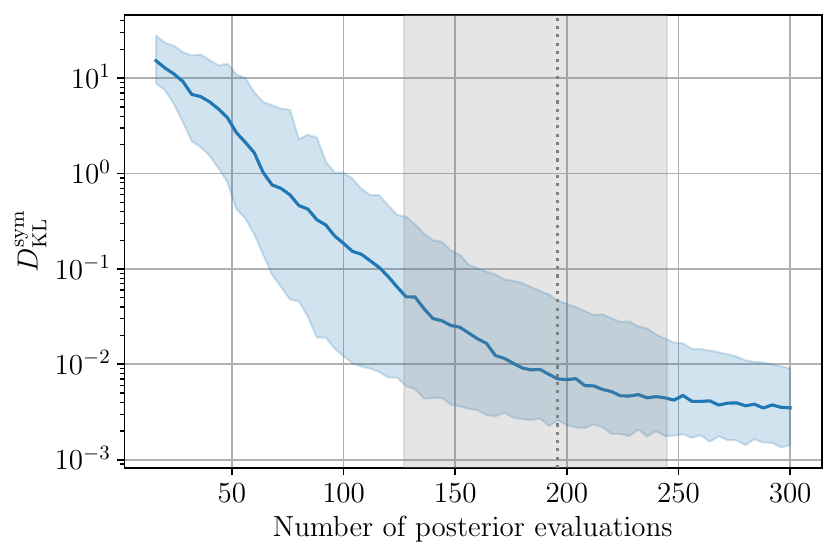}
    \caption{2d and 1d posterior distributions of a typical four-dimensional log-gaussian distribution (left) at convergence (180 posterior evaluations), and convergence with respect to the true model against number of accepted steps for 200 realizations, where the blue band shows the $\{25,50,75\}\%$-quantiles for the KL-divergence, and the grey band does the same for convergence as defined by the \texttt{CorrectCounter} criterion. (right). The posterior distribution is cut off in $x_0$ and $x_1$, which is correctly captured by \texttt{GPry}.}
    \label{fig:4dloggaussian}
\end{figure}

To test whether our algorithm is robust with respect to these kind of likelihoods we drew randomly correlated 4-dimensional Gaussians according to  \Cref{eq:randomly_correlated_gaussians} where the first two dimensions $\{x_0, x_1\}$ are sampled in log-space. The performance of the algorithm in this case is shown in \Cref{fig:4dloggaussian}. We recover the correct posterior shape and manage to sample the posterior accurately with only around $200$ samples. An additional benefit of this test is that it shows that our algorithm is robust with respect to cases where the mode has a hard prior cutoff ($|x_i|<2$ in this example).

\subsubsection{Curved degeneracies}
\label{ssec:curved}

\begin{figure}[t]
    \centering
    \subfloat[\enquote{Banana}-shaped degeneracy]{
        \includegraphics[width=0.35\textwidth]{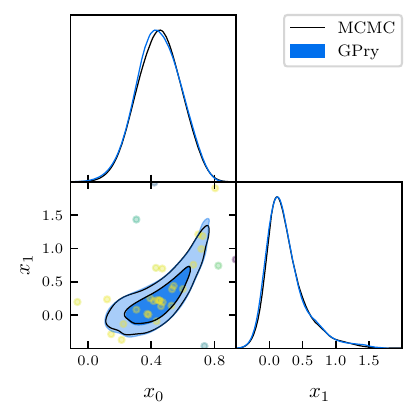}
        \includegraphics[width=0.45\textwidth]{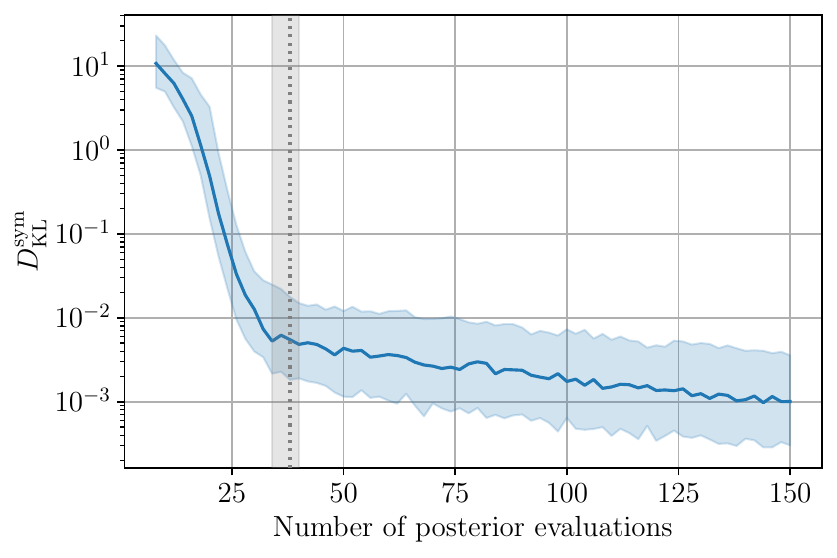}
        \label{fig:curved_degeneracy}
    }\\
    \subfloat[Rosenbrock likelihood]{
        \includegraphics[width=0.35\textwidth]{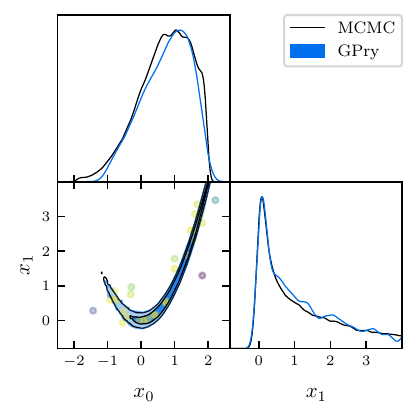}
        \includegraphics[width=0.45\textwidth]{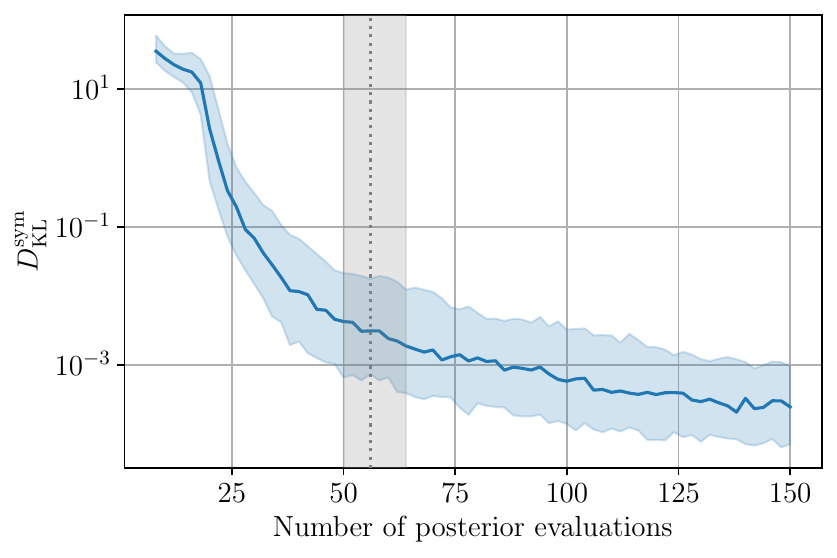}
        \label{fig:rosenbrock}
    }\\
    \subfloat[Gaussian ring]{
        \includegraphics[width=0.35\textwidth]{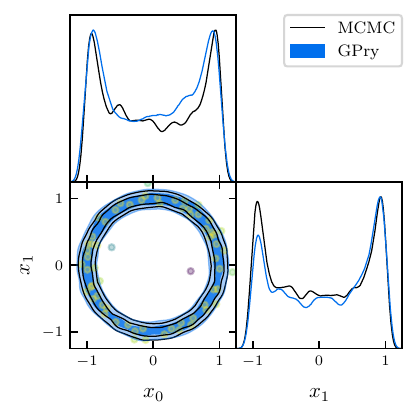}
        \includegraphics[width=0.45\textwidth]{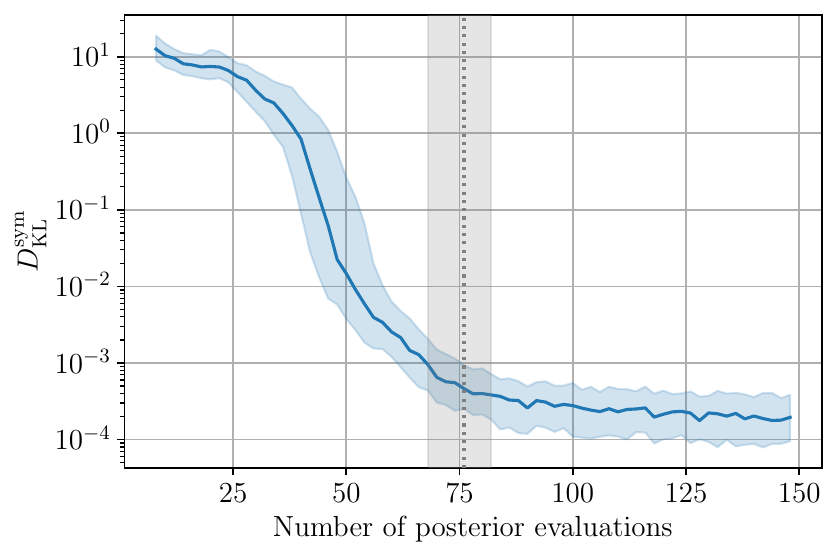}
        \label{fig:ring}
    }
    \caption{Performance tests for the non-Gaussian likelihoods with curved degeneracies presented in \Cref{ssec:curved}. For each case \textbf{Left:} 2d and 1d posterior distributions for typical converged realisations ($40$, $62$ and $68$ posterior evaluations, respectively); \textbf{Right:} convergence against number of accepted steps, where the blue and grey bands are defined as in \Cref{fig:4dloggaussian}. Even though these distributions display very non-Gaussian behaviours, their shape is correctly recovered without needing a large number of samples.\vspace*{-3\baselineskip}%Otherwise the figure alone doesn't even fit on a single page
    }
    \label{fig:curved_degeneracies}
\end{figure}

We also investigated whether more general curved degeneracies with different length-scales in the different parameter dimensions could be modeled correctly. For this, we use three examples. 
\begin{enumerate}
    \item Example one is a \enquote{banana}-shaped curved degeneracy, a slightly modified version of a benchmark found in \cite{curved_degeneracy}, which is based upon an eight-order polynomial in the exponent and exhibits a long tail in the $x_1 \approx 4 x_0^4$ direction. The log-likelihood of this distribution is
\begin{equation}
    \log \mathcal{L}(x_0, x_1) = -(10\cdot(0.45-x_0))^2/4 - (20\cdot(x_1/4-x_0^4))^2 \ .
\end{equation}
\Cref{fig:curved_degeneracy} shows how \texttt{GPry} performs at sampling this distribution. The posterior shape is correctly recovered (at around $\sim 40$ posterior evaluations) and shows good match with MCMC.
    \item Example two has a fourth-order polynomial in the exponent, but in this case the parameters are tuned in order to exhibit an extremely sharp cutoff away from the degeneracy direction and an extremely long tail along the degeneracy. This particularly pathological case is the Rosenbrock function, commonly used to test minimization algorithms. It is described by
    \begin{equation}
    \log \mathcal{L}(x_0, x_1) = -\frac{1}{2}\left[(a-x_0)^2+b(x_1-x_0^2)^2\right]~,
    \end{equation}
    where we set the parameters to their typical values of $a=1$ and $b=100$. It has a long, narrow, parabolic \enquote{ridge} along which the maximum lies. Since the parabolic degeneracy direction changes throughout, this is a good test for the robustness of \texttt{GPry} for distributions which do not show a clear axis of correlation or symmetry. We impose a uniform prior between $[-4,4]$ for both $x_0$ and $x_1$. The results for this posterior are displayed in \Cref{fig:rosenbrock}, which shows that even such a pathological posterior function can be accurately described by the \texttt{GPry} code, while still requiring a reasonably small number of posterior evaluations ($\sim 60$).
    \item The third example is a sharp ring-like posterior. The log-likelihood of this distribution is given by
    \begin{equation}
        \log \mathcal{L}(x_0, x_1) = -\frac{1}{2}\left[\frac{(\sqrt{x_0^2+x_1^2}-\mu)^2}{\sigma}+\log(2\pi\sigma^2)\right]~,
    \end{equation}
    with $\mu=1$ and $\sigma=0.05$. This produces a ring-shaped posterior distribution with the two very different scales $\mu$ (the location of the ring) and $\sigma$ (the width of the ring). Furthermore the maximum of this function is the ridge of the ring, making it especially hard to capture the full mode and sample the distribution correctly. Nevertheless our algorithm efficiently captures this mode within $\sim 75$ posterior evaluations and agrees well with MCMC. 
\end{enumerate} 

We note that for all of these non-Gaussian examples more posterior evaluations are required for convergence compared to the multi-variate Gaussian examples with the same dimensionalities. This is because the surrogate model requires more training samples to correctly capture the non-trivial shape and the extended tails.

\subsubsection{Multi-modal posteriors}\label{sssec:multi_modal_posteriors}

We also want to check the robustness of the \texttt{GPry} tool against mild multi-modality. For this, we make use of a modified Himmelblau function (which is commonly used in minimization studies). The log-posterior is defined as 
\begin{align}\label{eq:himmelblau}
    \log \mathcal{L}(x_0,x_1)=-\frac{1}{2}\left[a\cdot(x_0^2-x_1-11)^2+(x+y^2-7)^2\right]
\end{align}
where the term in the brackets corresponds to the Himmelblau function for $a=1$. We include this scaling factor $a$ in the first term in order to create a \enquote{mild} multi-modal posterior ($a=0.1$) with relatively connected modes which we compare to the full Himmelblau function ($a=1$).

\begin{figure}
    \centering
    \includegraphics[width=0.35\textwidth]{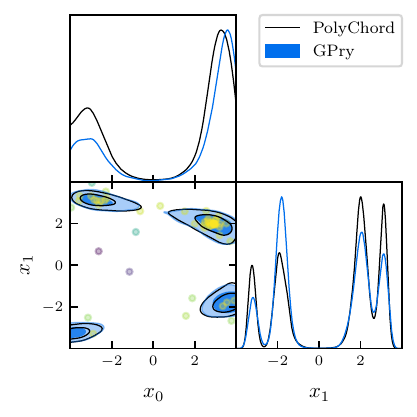}
    \includegraphics[width=0.45\textwidth]{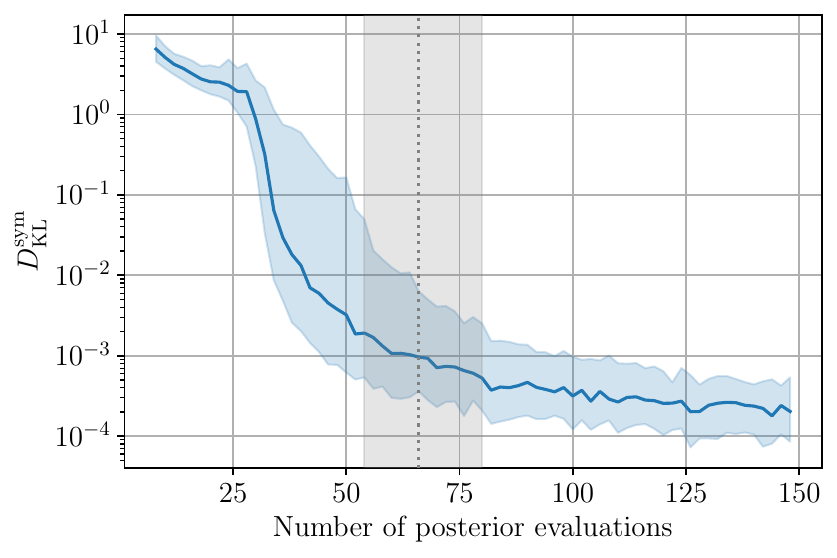}
    \caption{2d and 1d posterior distributions of a typical, converged realisation of the \enquote{mild} Himmelblau function (left) at convergence (58 posterior evaluations) and convergence against number of accepted steps (right), where the blue and grey bands are defined as in \Cref{fig:4dloggaussian}. The function has four modes which are all sampled but not weighed correctly by \texttt{GPry}. \texttt{GPry} on average needs few ($\lesssim 75$) samples to claim convergence.}
    \label{fig:himmelblau}
\end{figure}

\begin{figure}[h]
    \centering
    \includegraphics[width=0.45\textwidth]{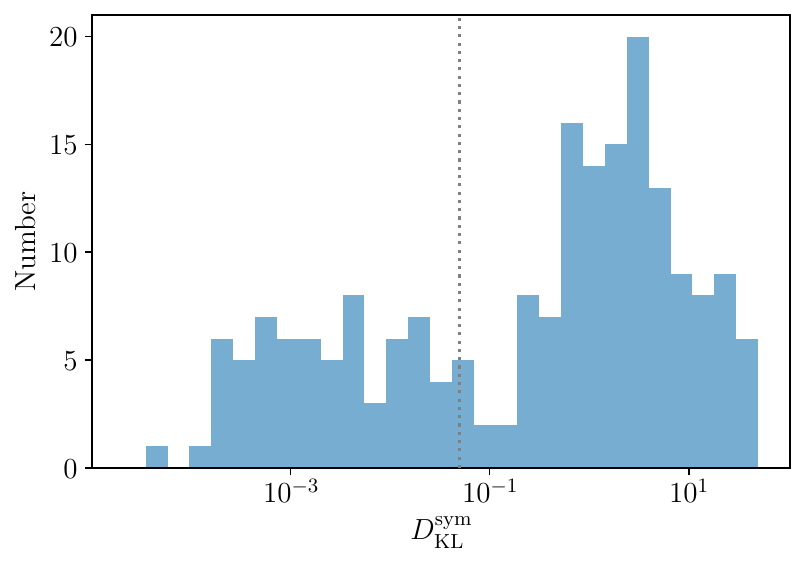}
    \includegraphics[width=0.45\textwidth]{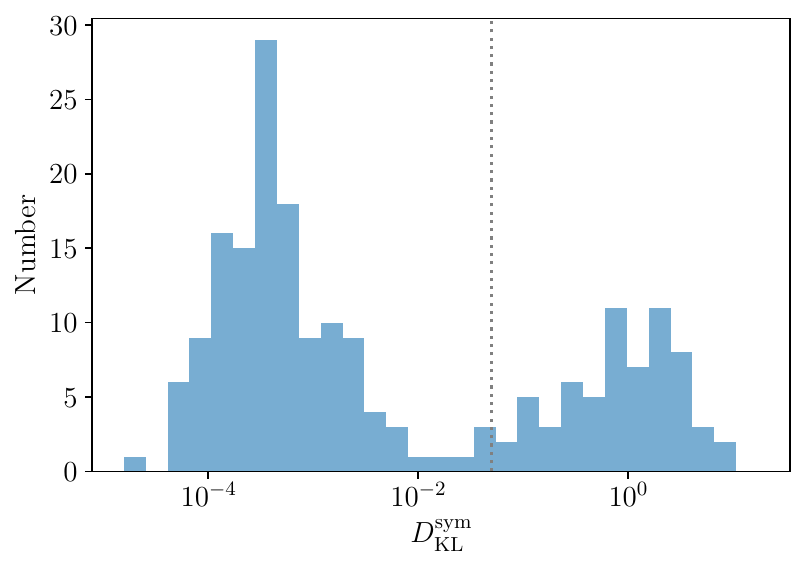}
    \caption{\textbf{Left:} Distribution of KL divergences at convergence according to \texttt{CorrectCounter} of the standard Himmelblau function. In many cases convergence is declared while not all of the four modes of the function have been explored, leading to large values of $D_\mathrm{KL}^\mathrm{sym}$. \textbf{Right:} Distribution of KL divergences for the same Himmelblau function at a budgeted, large number of samples (in this case 150). The distribution shows that sampling beyond nominal convergence of the \texttt{CorrectCounter} criterion would aid in improving the interpolation. Nonetheless, there still remain two modes: one at low values of $D^\mathrm{sym}_{\mathrm{KL}}$ (around $D^\mathrm{sym}_{\mathrm{KL}}=10^{-3}$) where all modes have been found and one at high values (around $D^\mathrm{sym}_{\mathrm{KL}}=1$) where some of the modes were not explored. Examples of this behaviour are shown in \Cref{fig:himmelblau_full_examples}.}
    \label{fig:himmelblau_full_hists}
\end{figure}

\begin{figure}[h]
    \centering
    \includegraphics[width=0.4\textwidth]{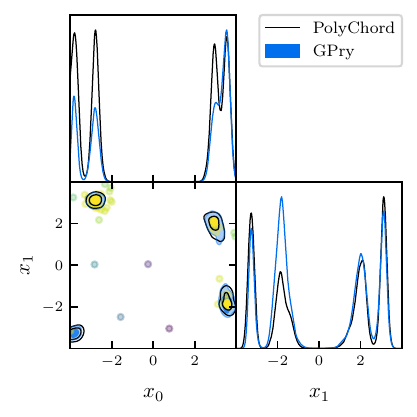}
    \includegraphics[width=0.4\textwidth]{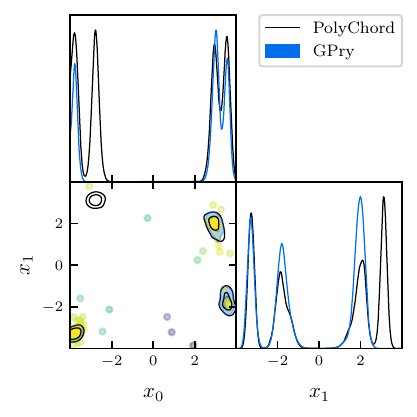}
    \caption{Exemplary 2d and 1d posterior distributions of the full Himmelblau function ($a=1$). \textbf{Left:} Contours of the algorithm finding all modes and converging at 102 posterior evaluations (although the 1D posteriors are not weighed correctly). \textbf{Right:} Example of the algorithm missing a mode completely and falsely claiming convergence. This problem arises when the posterior distribution to map has several disconnected modes. If one of the modes is missed completely early in the sampling procedure the GP surrogate and hence the acquisition procedure may deem this region irrelevant and not sample there. This behaviour is especially severe when the SVM classifies the region which contains the additional mode(s) as infinite.}
    \label{fig:himmelblau_full_examples}
\end{figure}

We show the results of sampling this distribution in \Cref{fig:himmelblau,fig:himmelblau_full_examples,fig:himmelblau_full_hists}. We observe that many runs do not correctly capture the modes. In general, we can distinguish three modes of failure of the \texttt{GPry} algorithm, nicely demonstrated in these examples.

\begin{enumerate}
    \item The algorithm can find and sample all modes, but not weigh them correctly in relation to each other. This is clearly visible in \Cref{fig:himmelblau}, where all modes are reliably sampled, but the 1D posterior reveals the incorrect weighting.
    \item The \texttt{CorrectCounter} criterion may falsely claim convergence and stop the sampling when some of the modes have been well explored, while further sampling might have revealed modes that have not been mapped. This is shown in \Cref{fig:himmelblau_full_hists}, where we compare the convergence to the true distribution (through the  $D_\mathrm{KL}^\mathrm{sym}$) when the \texttt{CorrectCounter} criterion has claimed convergence, with that of the runs at a larger number of samples (150 in this case). We observe that if the sampling had continued further, they would have been able to better map the underlying modes. See also \Cref{fig:himmelblau_full_examples} for two examples of these first two failure modes for the $a=1$ case. 
    \item The SVM could characterize a whole region as irrelevant due to a very deep intermediate valley even though a mode is present there. In that case, no amount of additional sampling would reveal the hidden mode. This failure mode does not occur for the $a=0.1$ or $a=1$ cases as the valleys are not deep enough there to be characterized as irrelevant.
\end{enumerate}

As such, we would like to stress that this package was designed with a focus on uni-modal distributions and that there is no guarantee that \emph{in general} all modes are captured or weighed correctly by the algorithm. Deeper investigations into multi-modal GP algorithms are left for future work. Note that for this distribution we used the PolyChord nested sampler \cite{polychord_1, polychord_2} for generating our reference contours and MC samples of the GP surrogate as it - unlike MCMC - reliably finds and explores all modes.

\subsection{Cosmology}

\begin{figure}
    \centering
    \includegraphics[width=0.45\textwidth]{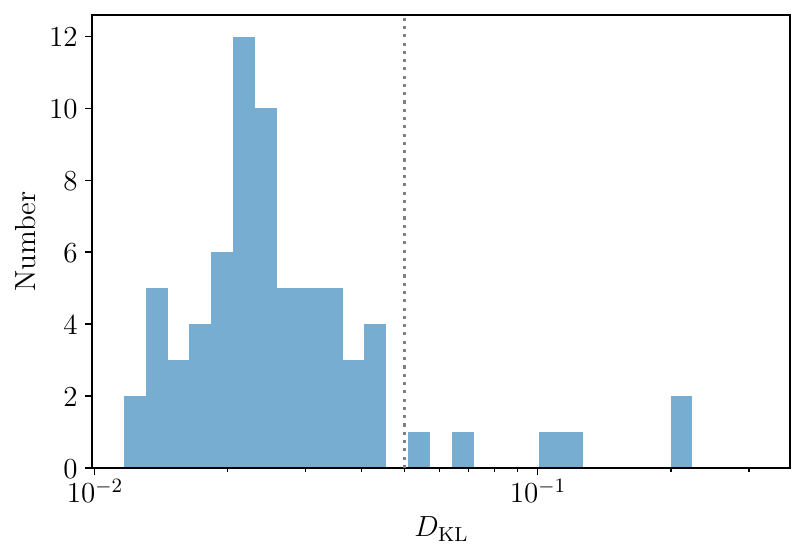}
    \includegraphics[width=0.45\textwidth]{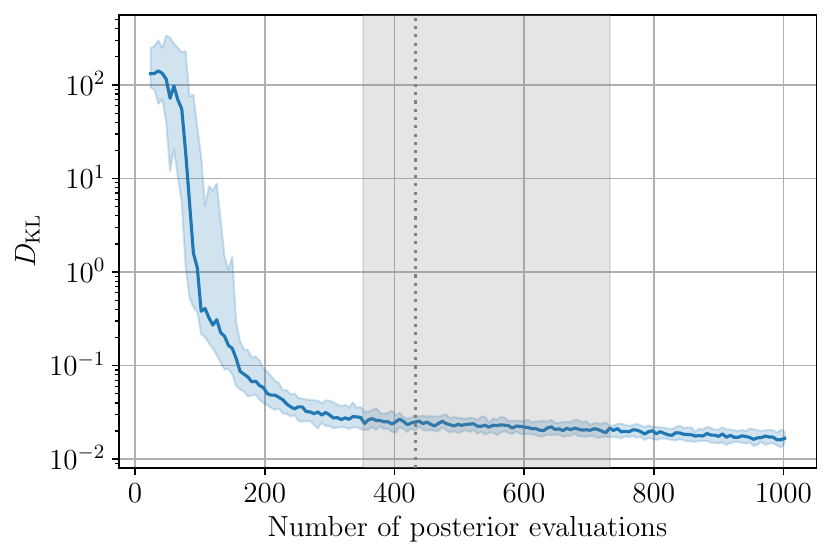}\\
    \includegraphics[width=0.75\textwidth]{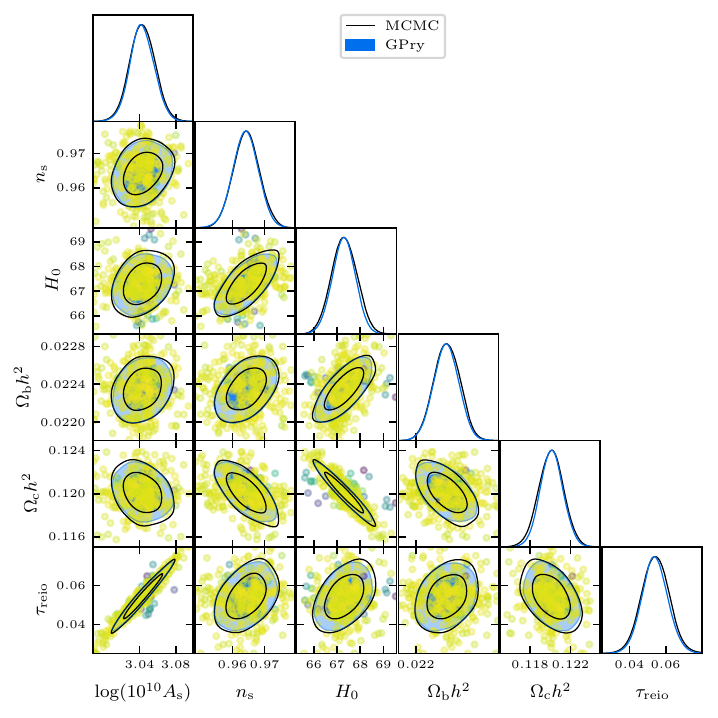}
    \caption{Constraints and convergence statistics in a $\Lambda$CDM model from Planck 2018 (TT,TE,EE,lensing) using the nuisance-marginalised Planck Lite likelihood. The given constraints could be obtained sampling only around $\sim 500$ (in this case 420) evaluations of the underlying theory code and likelihood. \textbf{Top:} Convergence statistics. \textbf{Bottom:} 1D posteriors and (68.3\%,95.4\%) contours of the 2D posteriors.}
    \label{fig:planck_lite_triangle}
\end{figure}

We also test the \texttt{GPry} tool in the context of cosmological applications, such as the inference of the posterior for Planck CMB anisotropy measurements (using the nuisance-marginalised Planck Lite likelihood of \cite{planck_lite_1,planck_lite_4} in the context of the 6-dimensional $\Lambda$CDM model). We performed $75$ separate runs of the \texttt{GPry} tool, converging on average within only around $500$ evaluations of the underlying theory code\footnote{Here we happen to be using \texttt{CLASS} \cite{class_1}, but since the \texttt{GPry} tool is fully interfaced with \texttt{Cobaya} \cite{cobaya}, other theory codes can be used as well.} The convergence history as well as the final KL-divergence upon termination through the convergence criterion are shown in \Cref{fig:planck_lite_triangle}. An exemplary case (close to the median in terms of required number of samples) is also shown in \Cref{fig:planck_lite_triangle}, where we can see that the constraints are very well aligned with those of the true posterior.

We note that the full Planck likelihood (including nuisance parameters) can also be modeled with \texttt{GPry}, but in this case the high dimensionality of the parameter space (27 dimensions in our case) makes the proposal of new points to start the acquisition optimization from (see Line~\ref{alg:acq_opt} of Algorithm~\ref{main_algoirthm}) rather difficult. If one uses the bestfit and covariance matrix of the Planck chains to propose these points instead, the acquisition function can be well optimized and the run does correctly map the posterior. We leave investigations of reaching convergence for the full Planck likelihood without any kind of a priori information (such as covariance matrix or bestfit) for future work.

\begin{figure}
    \centering
    \includegraphics[width=0.75\textwidth]{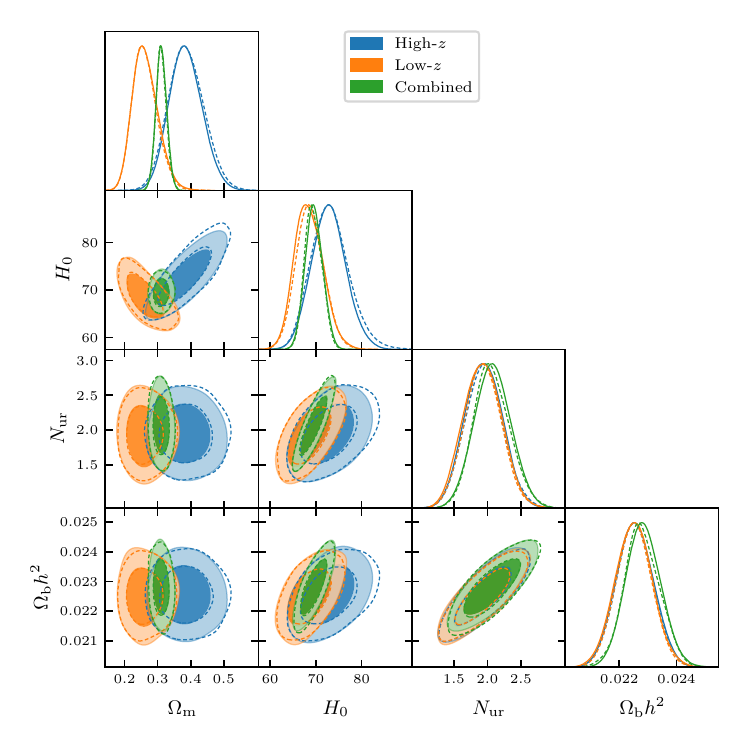}
    \caption{Triangle plot showing the marginalised constraints of the four-dimensional likelihood of BBN+BAO measurements for high-$z$, low-$z$ and combined likelihoods. \texttt{GPry} is able to recover all contours correctly with only very few (124, 108, 80) posterior evaluations. The contours that we recover are in excellent agreement with the constraints from MCMC.}
    \label{fig:bbn_bao_triangle}
\end{figure}

Another illustrative example is that of the combined Big Bang Nucleosynthesis (BBN) and Baryon Acoustic Oscillatons (BAO) measurements. We combine low redshift BAO from 6dFGS galaxies, the DR7 main galaxy sample, DR12 luminous red galaxies (together low-$z$), as well as high redshift BAO from DR16 quasars, DR16 Lyman-$\alpha$ based BAO, and their cross-correlations (together high-$z$), in order to constrain the Hubble constant and the matter composition of the Universe. The BBN likelihood we adopt is the same as in \cite{bbn_paper_nils}. For this case we vary the number of effective neutrinos $N_{\mathrm{eff}}$\,, corresponding to the addition of dark radiation to the $\Lambda$CDM model. This results in three possible data likelihood combinations, depending on whether we combine with the BBN data the the low redshift galaxy based BAO likelihoods (\enquote{low-$z$}), with the higher redshift Lyman-$\alpha$ and quasar BAO likelihoods (\enquote{high-$z$}), or we use both redshift samples (\enquote{combined}). For every combination, we sample the four-dimensional posterior using both MCMC and \texttt{GPry}. In \Cref{fig:bbn_bao_triangle} we show the resulting triangle plot which is in excellent agreement, demonstrating again the flexibility of GPry even when the underlying model or the used data sets are varied. The contours can be recovered with only around $\sim 100$ posterior evaluations (as opposed to the $\sim 10^4$ points used for the MCMC chains).

%%%%%%%%%%%%%%%%%%%%%%%%%%%%%%%%%%%%%%%%%%%%%%%%%%%%%%%%%%%%%%%%%%%%%%%%%%%%%%%%
\section{Conclusions} \label{sec:conclusions}
In this paper we presented the \texttt{GPry} algorithm and Python package implementation. As shown with both synthetic and cosmological likelihoods our algorithm requires vastly less posterior evaluations for generating a fair Monte Carlo sample for Bayesian Inference than current state-of-the-art MCMC and nested samplers. We report up to multiple orders of magnitude improvements in the number of posterior evaluations required, as well as in wall-clock computation time savings, making this algorithm very promising for slow likelihood codes. This not only speeds up inference significantly but also reduces its carbon footprint. Furthermore, we open a new window of possibilities by enabling inference from extremely slow likelihoods ($\gtrsim\mathrm{minutes}$ per evaluation), which otherwise would be impossible to sample, since traditional samplers might take months to converge. In addition,  since our algorithm does not rely on specialized hardware (such as GPUs) or any kind of pre-training, it can be used as a drop-in replacement for traditional Monte Carlo samplers. Particularly in the case of cosmological applications it benefits from an interface with \texttt{Cobaya}.

Despite the algorithm's impressive performance, there is still ample room for improvement both in terms of speed and robustness. In a future series of papers we plan to explore four main avenue: (i) The overhead of constructing the GP surrogate model and the acquisition procedure could be further minimized by using clever numerical techniques, allowing \texttt{GPry} to outcompete traditional MC samplers even for fast likelihoods. (ii) As discussed in \Cref{sssec:multi_modal_posteriors} \texttt{GPry} currently is optimized for unimodal posterior distributions; it would be desirable to increase the robustness towards strongly multi-modal posteriors by generating the starting points for the acquisition optimization in a special way, and using clustering algorithms to track different modes separately. (iii) For likelihood distributions with significant stochastic or numerical noise, it would be beneficial to automatically adapt the noise term in \cref{joint_distribution} without requiring prior knowledge. (iv) For high dimensionalities the current methods of proposing additional points for restarting hyperparameter optimization and sample acquisition are still relatively naive. Similarly, the overhead of the underlying operations performed on the GP increases strongly with the number of acquired samples. Both of these hurdles can be overcome with novel approaches, potentially unlocking even the regime of high-dimensional likelihoods for further optimization with \texttt{GPry}.

The \texttt{GPry} algorithm and python package presented in this work enables parameter inference in cosmology without high computational and environmental costs. This opens up new possibilities for Bayesian inference on costly likelihood functions which have been computationally unfeasible before. With many avenues of optimization of the code-base and algorithm still left to explore, \texttt{GPry} will only continue to improve in efficiency and accuracy.

\subsubsection*{Acknowledgements}

We thank Julien Lesgourgues, Antony Lewis, Andrew Liddle and Marcos Pellejero Ib\'a\~nez for useful discussions. This project was initiated when all authors were working at the TTK institute of RWTH Aachen University. We also acknowledge the use of the JARA computing cluster of the RWTH Aachen University under project \texttt{jara0184}. N.~S.~acknowledges support from the Maria de Maetzu fellowship grant: CEX2019-000918-M, financiado por MCIN/AEI/10.13039/501100011033. J.~E. acknowledges support by the ROMFORSK grant project no.~302640.

%%%%%%%%%%%%%%%%%%%%%%%%%%%%%%%%%%%%%%%%%%%%%%%%%%%%%%%%%%%%%%%%%%%%%%%%%%%%%%%%

\appendix

\section{Posterior scale in higher dimensions}\label{app:threshold}

When considering a problem with a larger number of dimensions, there are a few aspects of the problem that require special care. It is a well-known fact that for a 1-dimensional Gaussian the region defined by one standard deviation around the mean contains $\approx 68\%$ of the total probability mass. The generalisation to higher dimensionality is non-trivial: for multivariate Gaussians, considering distances defined in units of the covariance matrix (\textit{Mahalanobis distance}), the region defined by a unit away from the mean contains a smaller and smaller fraction of the total probability mass as dimensionality goes up. This is, of course, nothing more than the \emph{curse of dimensionality}, and it will be present in most of the inference problems that we target in this study.\footnote{It affects distributions with significant tails, and most well-behaved distributions show tails, including those in the exponential family (which includes multivariate Gaussians).}

In our context of modelling a probability density function, this is reflected in the dynamic range of log-probability that needs to be carefully modelled, meaning that given some confidence limit (\CL{}) up to which we want our model to be especially precise, the difference between the log-posterior corresponding to that \CL{} and the maximum log-posterior will depend on dimensionality. This dynamic range will show up at three different steps of the algorithm, explicitly in the treatment of infinities and extreme values in \Cref{ssec:svm} and the convergence criterion in \Cref{ssec:convergence}, and implicitly in the choice of the acquisition hyperparameter in \Cref{ssec:zeta}. Taking into account this dimensionality scaling in the ways explained below has proven to dramatically improve the performance of our algorithm.

In order to give a rough order-of-magnitude estimate for this log-posterior scaling, we can turn towards a multivariate Gaussian distribution of the same dimensionality. Treated as a random variable itself, a multivariate Gaussian log-probability is proportional to the sum of $N_d$ independent standard 1-dimensional Gaussian random variables (up to a linear covariance-diagonalizing transformation). Thus the value of the Gaussian log-posterior when multiplied by $-2$ follows a $\chi^2$ distribution with $N_d$ degrees of freedom. Defining $\Delta \chi^2 = 2 [\max(\log p)-\log p]$ we find $\Delta \chi^2 \sim {\chi^2}_{N_d}$\,. 

We can use this to compute the posterior range corresponding to different \CL{}s defined by the posterior mass $\epsilon$ that they leave out, using the $\chi^2$ cumulative distribution function $F_{N_d}$ (where $N_d$ is the number of degrees of freedom):
\begin{equation}
  1 - \varepsilon = F_{N_d}(\Delta \chi^2)\,.
\end{equation}
When referring to \CL{}s in higher dimensions, we can alternatively name them as their 1D equivalent normal Gaussian extent ($1-\epsilon=0.683$ for 1-$\sigma$, $1-\epsilon=0.954$ for 2-$\sigma$, etc.). As such, in the following when we refer to a $n$-$\sigma$ contour in an arbitrary dimensionality within this paper, we explicitly refer to the \CL{} corresponding to that number of standard deviations in a $1D$ Gaussian. 
Explicitly, since $F_1(x) = \mathrm{erf}(\sqrt{x/2})$, and given that the value of a $\chi^2_1$ random variable represents the squared number of standard deviations away from the mean in the corresponding Gaussian, we can simply write $1 - \varepsilon = \mathrm{erf}(n/\sqrt{2})$ for a given n-$\sigma$ \CL{}.

With this, we can get the expected scaling in $N_d$ dimensions corresponding to a n-$\sigma$ probability mass as
\begin{equation}
    [\Delta \chi^2] (n) =F^{-1}_{N_d}\left[\mathrm{erf}(n/\sqrt{2})\right]\,.
 \end{equation}
As an example, the 2-$\sigma$ ($1-\epsilon=0.954$) contour corresponds to a range $[\Delta \chi^2](2) = 9.72$ in 4 dimensions and $[\Delta \chi^2](2) = 15.79$ in 8 dimensions.

In \Cref{ssec:svm} we have used this result to derive the threshold value $T$ for the SVM (the criterion for deciding if a sample has a sufficient log-posterior to be added to the GP) by imposing $T= [\Delta \chi^2](n_T)/2$ which is the scaling of the log-posterior for $n_T=20$ ($\epsilon \approx 5.5 \cdot 10^{-89}$), and has been found to work well for most practical applications (including all examples in this work). This prescription ensures dimensional consistency: choices of $T$ as absolute values do not work well across different dimensions, causing the SVM to be too permissive in low dimensions (does not capture extreme values efficiently) and too stringent in high dimensions (points with significant log-posterior are excluded).

We have also used this result to scale the tolerance of the convergence criterion in \Cref{ssec:convergence}. In particular, since the absolute threshold $\epsilon_\mathrm{abs}$ is compared against differences in absolute values of $\log p$, we are scaling it as these differences do for a fixed difference in credibility, in particular that of the first $\sigma$ credible (hyper)volume: $\epsilon_\mathrm{abs} = 0.01 [\Delta \chi^2](1)$.

\begin{figure}[t]
    \centering
    \includegraphics[width=1\textwidth]{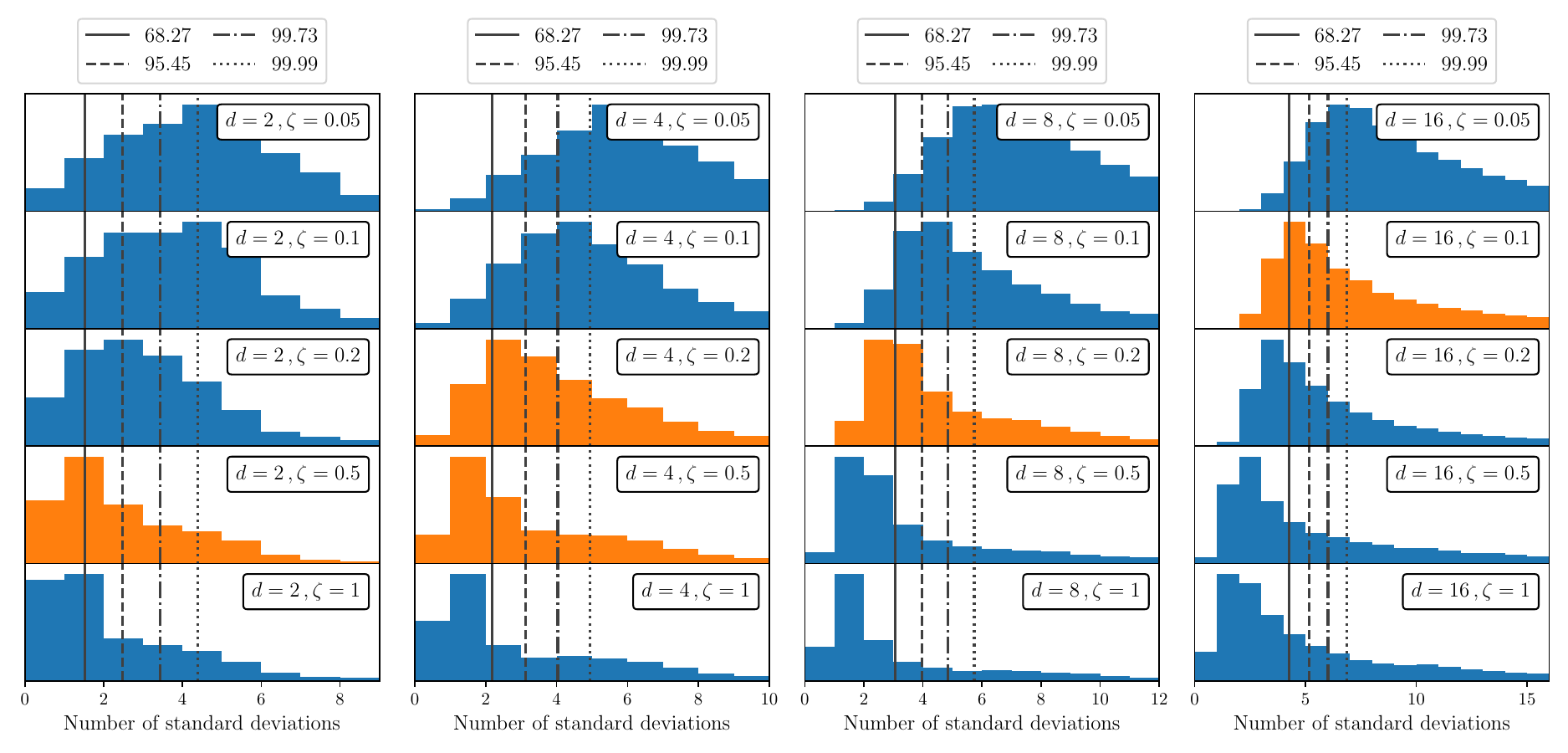}
    \caption{
    Histograms of aggregated Mahalanobis distances of the points in the training sets of the realisations used in \Cref{fig:zeta_vs_kl}, for different dimensionalities (columns) and values of $\zeta$ (rows). The optimal $\zeta$'s from the experimental relation $\zeta = N_d^{-0.85}$ are highlighted in orange/clear (for $d=4$, the two closest values are both highlighted). A remarkable result is that efficiency at converging (the criterion imposed to get the optimal $\zeta$'s) is maximised when the distribution of training points are centered around the same CL ($68\%$) in all dimensionalities, likely imposed indirectly by using the dimensionally-consistent KL divergence to assess convergence when selecting optimal $\zeta$'s.
    }
    \label{fig:zeta_grid}
\end{figure}

Regarding the dimensionality scaling of the learning hyperparameter, it is not trivial to find an analytic prescription to write $\zeta$ as a function of $[\Delta \chi^2](n)$. As discussed in \Cref{ssec:zeta} we have derived an experimental scaling $\zeta = N_d^{-0.85}$, which corresponds to the value of $\zeta$ that leads to convergence in the smallest number of posterior evaluations, as demonstrated in \Cref{fig:zeta_vs_kl}. We can check a posteriori how these optimal dimensional-dependent $\zeta$'s relate to the CL's at these dimensionalities. To do that, we compute the Mahalanobis distances of all points in the training sets of all the realisations used for \Cref{fig:zeta_vs_kl}, and create histograms of these distances for each dimensionality and $\zeta$ in \Cref{fig:zeta_grid}. In this figure, we highlight the cases that converged most efficiently in orange/clear, as assessed by the dimensionally-consistent Kullback-Leibler divergence. We observe that convergence is achieved more efficiently when $\zeta$ is such that the distribution of training points is centered around the same CL ($68\%$) in all dimensionalities. This underlines the idea that the dimensionally-dependent CL’s should set the relevant scales in the surrogate model for optimal efficiency, and is possibly in fact a consequence of having used a dimensionally-consistent method (the Kullback-Leibler divergence) to assess convergence.

%%%%%%%%%%%%%%%%%%%%%%%%%%%%%%%%%%%%%%%%%%%%%%%%%%%%%%%%%%%%%%%%%%%%%%%%%%%%%%%%

\section{KL divergence}\label{app:kl}

A natural way of assessing the \enquote{similarity} of two distributions is the Kullback-Leibler (KL) divergence. The KL divergence between two continuous probability distributions $P$ and $Q$ with probability density functions $p(\ve{x})$ and $q(\ve{x})$ is defined as \cite{kl_divergence}
\begin{align}
    D_{\mathrm{KL}}(P||Q) = \int p(\ve{x})\log\left(\frac{p(\ve{x})}{q(\ve{x})}\right)\, \d \ve{x}~.
\end{align}
Furthermore there exists a symmetric version (often called Jeffreys divergence) 
\begin{align}
    D_{\mathrm{KL}}^{\mathrm{sym}}(P, Q) = \frac{1}{2}\left(D_{\mathrm{KL}}(P||Q) + D_{\mathrm{KL}}(Q||P)\right)
\end{align}
A smaller value means that the two posteriors are in better agreement, and one typically wants $D_{\mathrm{KL}}(P||Q) \ll 1$ for good agreement. The dimensionality consistency of the KL divergence guarantees that a given value for the divergence characterizes similar differences across dimensionalities.

To compute the KL divergence explicitly, one can use the fact that the points in a Monte Carlo sample of $P$ are distributed as $p(\ve{x})\d \ve{x}$. One can thus approximate the integral as a sum of the quantity $\log p(\ve{x}_i)-\log q(\ve{x}_i)$ over all points in the MC sample (multiplied by their respective weights/multiplicities).

We can use the KL divergence to assess the convergence towards the true distribution of a GP surrogate model, if a sample from the true distribution can be obtained with the usual MC methods (e.g.\ in the test cases presented in \Cref{sec:examples}). In that case, $\log p(\ve{x}_i)$ would be the true log-posterior at point $i$ in the MC sample, and $\log q(\ve{x}_i)$ would be the emulated log-posterior from \texttt{GPry} at that same point. In practical applications where an MC sample of the true posterior is not possible to obtain, the KL divergence can be used in a similar fashion to define a convergence criterion by comparing GP surrogate models at consecutive iterations of the \texttt{GPry} algorithm, summing over an MC sample of the GP surrogate model at a particular step (see \Cref{ssec:convergence}).

In order to save a significant amount of memory, when using the KL divergence for the purpose of a convergence criterion, instead of integrating a full MCMC, we only store the information from the mean and the covariance matrix. This is equivalent to approximating the underlying distributions as multi-variate Gaussian distributions (with mean $\mathfrak{m}$ and covariance $\ve{C}$). While this is a bad description for the distribution itself, it is often the case that when the multi-variate Gaussian approximation of a distribution agrees to a high level of precision with that of another distribution, so do the underlying distributions. Under this approximation the KL divergence is simply given by 
\begin{align}\label{eq:kl_gaussian}
    D_{\mathrm{KL}}(P||Q) \approx \frac{1}{2}\left(\mathrm{tr}\left(\ve{C}_Q^{-1}\ve{C}_P\right)-d+(\mathfrak{m}_Q-\mathfrak{m}_P)^T\ve{C}_Q^{-1}(\mathfrak{m}_Q-\mathfrak{m}_P)+\log\left(\frac{\mathrm{det}\ve{C}_Q}{\mathrm{det}\ve{C}_P}\right)\right)~.
\end{align}

Whether using the MC-summed or the Gaussian approximation for the KL divergence, using it to naively define a convergence criterion, can be problematic, since running a full Monte Carlo sample at every acquired point, or at every iteration, would dominate the overhead of the algorithm. To reduce this computational cost, we take a number of decisions: Before deciding whether to re-run the Monte Carlo sample, we reweigh the previous one and compute the KL divergence between it and the previous estimate. We then re-use the reweighed one if the KL divergence between original and reweighed is small enough. We also relax the convergence criterion of the Monte Carlo algorithm early in the sampling procedure as convergence there is rather unlikely, so we do not need a high-quality estimation of the mean and covariance at that point.

Convergence is then determined by defining a threshold $c$ value such that the algorithm stops when $D_{\mathrm{KL}} < c$ during $n$ iterations, suggesting that the interpolation of the posterior distribution has stabilised. We set $n=2$ as the default.

Even with these improvements this method still produces considerable computational overhead, mainly due to the fact that running a Monte Carlo chain needs a large number of samples from the GP, especially as the number of dimensions increases.

\bibliographystyle{JHEP}
\bibliography{bibliography}

\end{document}